\documentclass[11pt]{article}
\pdfoutput=1
\usepackage[T1]{fontenc}
\usepackage{lmodern}
\usepackage[protrusion=true,expansion=true]{microtype}
\usepackage{amsmath,amssymb,amsfonts,amsthm}
\usepackage{subcaption}
\usepackage{graphicx}
\usepackage{setspace}
\usepackage{fullpage}
\usepackage[backref=page]{hyperref}
\usepackage{color}
\usepackage{wrapfig}
\usepackage{tikz}
\usetikzlibrary{decorations.pathreplacing}
\usepackage{algorithm}
\usepackage[noend]{algpseudocode}
\usepackage[framemethod=tikz]{mdframed}
\usepackage{xspace}
\usepackage{pgfplots}
\usepackage{framed}
\usepackage{thmtools}
\usepackage{thm-restate}
\usepackage{tabu}
\usepackage{fancyhdr}
\usepackage{comment}
\pgfplotsset{compat=1.5}

\newtheorem{theorem}{Theorem}[section]

\newtheorem{lemma}[theorem]{Lemma}

\newtheorem{definition}[theorem]{Definition}

\newtheorem{framework}[theorem]{Framework}

\newtheorem{fact}[theorem]{Fact}

\newenvironment{proofof}[1]{\begin{trivlist} \item {\bf Proof
#1:~~}}
  {\qed\end{trivlist}}

\newcommand{\namedref}[2]{\hyperref[#2]{#1~\ref*{#2}}}
\newcommand{\thmlab}[1]{\label{thm:#1}}
\newcommand{\thmref}[1]{\namedref{Theorem}{thm:#1}}
\newcommand{\lemlab}[1]{\label{lem:#1}}
\newcommand{\lemref}[1]{\namedref{Lemma}{lem:#1}}

\newcommand{\seclab}[1]{\label{sec:#1}}
\newcommand{\secref}[1]{\namedref{Section}{sec:#1}}
\newcommand{\applab}[1]{\label{app:#1}}
\newcommand{\appref}[1]{\namedref{Appendix}{app:#1}}
\newcommand{\factlab}[1]{\label{fact:#1}}
\newcommand{\factref}[1]{\namedref{Fact}{fact:#1}}

\newcommand{\figlab}[1]{\label{fig:#1}}
\newcommand{\figref}[1]{\namedref{Figure}{fig:#1}}
\newcommand{\alglab}[1]{\label{alg:#1}}
\renewcommand{\algref}[1]{\namedref{Algorithm}{alg:#1}}
\newcommand{\tablelab}[1]{\label{tab:#1}}
\newcommand{\tableref}[1]{\namedref{Table}{tab:#1}}
\newcommand{\deflab}[1]{\label{def:#1}}
\newcommand{\defref}[1]{\namedref{Definition}{def:#1}}

\newcommand{\frameref}[1]{\namedref{Framework}{frame:#1}}
\newcommand{\framelab}[1]{\label{frame:#1}}

\newcommand{\PPr}[1]{\ensuremath{\mathbf{Pr}\left[#1\right]}}

\newcommand{\Ex}[1]{\ensuremath{\mathbb{E}\left[#1\right]}}

\renewcommand{\O}[1]{\ensuremath{\mathcal{O}\left(#1\right)}}
\newcommand{\tO}[1]{\ensuremath{\tilde{\mathcal{O}}\left(#1\right)}}
\newcommand{\eps}{\varepsilon}

\def \dist    {\mdef{\text{dist}}}

\def \ba    {\mdef{\mathbf{a}}}

\def \calC    {\mdef{\mathcal{C}}}

\def \calE    {\mdef{\mathcal{E}}}

\def \calK    {\mdef{\mathcal{K}}}

\def \calO    {\mdef{\mathcal{O}}}

\def \calQ    {\mdef{\mathcal{Q}}}

\def \calS    {\mdef{\mathcal{S}}}

\def \bA    {\mdef{\mathbf{A}}}
\def \bB    {\mdef{\mathbf{B}}}
\def \bD    {\mdef{\mathbf{D}}}
\def \bE    {\mdef{\mathbf{E}}}
\def \bF    {\mdef{\mathbf{F}}}
\def \bG    {\mdef{\mathbf{G}}}
\def \bH    {\mdef{\mathbf{H}}}
\def \bI    {\mdef{\mathbb{I}}}

\def \bM    {\mdef{\mathbf{M}}}
\def \bP    {\mdef{\mathbf{P}}}
\def \bR    {\mdef{\mathbf{R}}}

\def \bT    {\mdef{\mathbf{T}}}
\def \bQ    {\mdef{\mathbf{Q}}}
\def \bV    {\mdef{\mathbf{V}}}
\def \bU    {\mdef{\mathbf{U}}}

\def \bY    {\mdef{\mathbf{Y}}}
\def \bZ    {\mdef{\mathbf{Z}}}
\def \bW    {\mdef{\mathbf{W}}}
\def \bb    {\mdef{\mathbf{b}}}

\def \br    {\mdef{\mathbf{r}}}

\def \bv    {\mdef{\mathbf{v}}}
\def \bx    {\mdef{\mathbf{x}}}
\def \by    {\mdef{\mathbf{y}}}
\def \bz    {\mdef{\mathbf{z}}}

\newcommand{\mdef}[1]{{\ensuremath{#1}}\xspace}  

\DeclareMathOperator*{\polylog}{polylog}
\DeclareMathOperator*{\poly}{poly}

\DeclareMathOperator*{\Cost}{cost}

\newcommand{\ignore}[1]{}

\newif\ifnotes\notestrue 
\ifnotes
\newcommand{\david}[1]{\textcolor{blue}{{\bf (David:} {#1}{\bf ) }} \marginpar{\tiny\bf
             \begin{minipage}[t]{0.5in}
               \raggedright S:
            \end{minipage}}}
\newcommand{\samson}[1]{\textcolor{red}{{\bf (Samson:} {#1}{\bf ) }} \marginpar{\tiny\bf
             \begin{minipage}[t]{0.5in}
               \raggedright S:
            \end{minipage}}}

\newcommand{\vincent}[1]{\textcolor{purple}{{\bf (Vincent:} {#1}{\bf ) }} \marginpar{\tiny\bf
             \begin{minipage}[t]{0.5in}
               \raggedright V:
            \end{minipage}}} 
\newcommand{\liudeng}[1]{\textcolor{magenta}{{\bf (Liudeng:} {#1}{\bf ) }} \marginpar{\tiny\bf
             \begin{minipage}[t]{0.5in}
               \raggedright L:
            \end{minipage}}}
\else
\newcommand{\samson}[1]{}
\newcommand{\liudeng}[1]{}
\newcommand{\david}[1]{}
\newcommand{\vincent}[1]{}
\fi

\makeatletter
\renewcommand*{\@fnsymbol}[1]{\textcolor{mahogany}{\ensuremath{\ifcase#1\or *\or \dagger\or \ddagger\or
 \mathsection\or \triangledown\or \mathparagraph\or \|\or **\or \dagger\dagger
   \or \ddagger\ddagger \else\@ctrerr\fi}}}
\makeatother

\providecommand{\email}[1]{\href{mailto:#1}{\nolinkurl{#1}\xspace}}

\definecolor{mahogany}{rgb}{0.75, 0.25, 0.0}
\definecolor{darkblue}{rgb}{0.0, 0.0, 0.55}
\definecolor{darkpastelgreen}{rgb}{0.01, 0.75, 0.24}
\definecolor{darkgreen}{rgb}{0.0, 0.2, 0.13}
\definecolor{darkgoldenrod}{rgb}{0.72, 0.53, 0.04}
\definecolor{darkred}{rgb}{0.55, 0.0, 0.0}
\definecolor{forestgreenweb}{rgb}{0.13, 0.55, 0.13}
\definecolor{greencss}{rgb}{0.0, 0.5, 0.0}
\definecolor{bleudefrance}{rgb}{0.19, 0.55, 0.91}
\definecolor{darkpastelpurple}{rgb}{0.59, 0.44, 0.84}
\definecolor{darkcerulean}{rgb}{0.03, 0.27, 0.49}

\hypersetup{
     colorlinks   = true,
     citecolor    = darkcerulean,
	 linkcolor	  = darkcerulean
}

\fancypagestyle{pg}
{
\lhead{}
\rhead{}
\cfoot{--\ \thepage\ --}

}

\AtBeginDocument{%
  \DeclareFontShape{T1}{lmr}{m}{scit}{<->ssub*lmr/m/scsl}{}%
}
\title{Space-Optimal Streaming Algorithms via Efficient Encodings}
\author{
Liudeng Wang\thanks{Texas A\&M University. E-mail: \email{eureka@tamu.edu}. The author is supported in part by NSF CCF-2335411.}
\and
David P. Woodruff\thanks{Carnegie Mellon University. E-mail: \email{dwoodruf@andrew.cmu.edu}. The author is supported in part by Office of Naval Research award number N000142112647 and a Simons Investigator Award.
}
\and
Shenghao Xie\thanks{Carnegie Mellon University. E-mail: \email{shenghax@andrew.cmu.edu}. The author is supported in part by NSF CCF-2335411, and in part by Office of Naval Research award number N000142112647 and a Simons Investigator Award. Significant part of the work was completed when the author was affiliated with Texas A\&M University.
}
\and
Samson Zhou\thanks{Texas A\&M University. E-mail: \email{samsonzhou@gmail.com}. The author is supported in part by NSF CCF-2335411, and gratefully acknowledges funding provided by the Oak Ridge Associated Universities (ORAU) Ralph E. Powe Junior Faculty Enhancement Award. \\
The work was initialized while David P. Woodruff and Samson Zhou were visiting the Institute for Emerging CORE Methods in Data Science (EnCORE) supported by the NSF grant 2217058. 
The work was conducted in part while David P. Woodruff and Samson Zhou were visiting the Simons Institute for the Theory of Computing as part of the Sublinear Algorithms program.
}
}
\date{}
\begin{document}
\pagestyle{pg}

\maketitle
\thispagestyle{empty}

\begin{abstract}
We study the problem of constructing coresets on data streams: Given a dataset of $n$ points in $\mathbb{R}^d$ arriving sequentially, the goal is to maintain a small weighted subset that accurately approximates the objective values of the entire dataset for a given class of query functions. Previous methods like merge-and-reduce and online sensitivity sampling required more space than the size of an optimal such subset in the offline setting. In this work, we show that such overheads are not necessary by presenting space-optimal streaming algorithms for central high-dimensional optimization problems such as Euclidean $(k,z)$-clustering and $L_p$ subspace embeddings, as well as applications to projection-cost preserving sketches. 

For $(k,z)$-clustering, our streaming algorithm uses memory independent of the number $n$ of input points (in words of memory) and the aspect ratio $\Delta$, yielding a coreset with an optimal $\tilde{\mathcal{O}}\left(\frac{dk}{\min(\varepsilon^4,\varepsilon^{z+2})}\right)$ words of memory for accuracy parameter $\varepsilon\in(0,1)$. For $L_p$ subspace embeddings, where the rows of matrix $\mathbf{A} \in \mathbb{R}^{n \times d}$ arrive in an insertion-only stream, we achieve an efficient algorithm that uses optimal space for different ranges of $p$: $\tilde{\mathcal{O}}\left(\frac{d^2}{\varepsilon^2}\right)$ words for $p\le 2$ and $\tilde{\mathcal{O}}\left(\frac{d^{p/2+1}}{\varepsilon^2}\right)$ words for $p>2$, both independent of the number of rows $n$. Thus, our work shows that streaming algorithms can match offline algorithms in space complexity. 
\end{abstract}

\section{Introduction}
A core challenge in large-scale machine learning systems is processing datasets that far exceed standard computer memory capacities.
To solve this issue, the \emph{streaming model} has emerged as a fundamental paradigm, in which elements of a dataset arrive sequentially in a data stream and an algorithm must process each arriving element immediately and irrevocably as it cannot revisit discarded data as the stream moves on. 
The primary goal is to design algorithms that can efficiently approximate, compute, or detect properties of the underlying dataset with minimal memory.
Thus, streaming algorithms would ideally excel at summarizing large-scale, evolving datasets under strict memory constraints. 
Despite significant progress, it remains open whether such fundamental tasks inherently require more space in the streaming model compared to their offline counterparts. 

\paragraph{Coresets.}
One of the most common data summarization approaches in streaming algorithms is to collect a \emph{coreset} \cite{Har-PeledM04,FeldmanL11}, which is a small, weighted subset that approximately preserves a target cost function.
Formally, given a set of queries $\calQ$ induced by a streaming problem $P$, a dataset $X$ of $n$ elements, and a cost function $\Cost(\cdot,\cdot)$ defined by the problem context, a (strong) $(1+\eps)$-coreset for $(X,P)$ is a weighted subset $Y$ of $X$ such that for all queries $q \in \calQ$, $\Cost(q,Y) \in (1\pm\eps) \cdot \Cost(q,X)$. 
The appeal of coresets lies in their ability to provide nearly identical cost values with reduced space complexity, enabling streaming algorithms to compute accurate approximations under memory constraints.

Perhaps the most important application of coresets is to Euclidean $(k,z)$-clustering, in which given a dataset $X=\{x_1,\ldots,x_n\} \subset \mathbb{R}^d$, a metric $\dist(\cdot, \cdot)$, a cluster parameter $k>0$, and a positive integer exponent $z>0$, and the goal is to minimize the clustering cost across all sets $\calC$ of at most $k$ centers: 
\begin{equation*}\min_{\calC\subset\mathbb{R}^d, |\calC|\le k}\Cost(X,\calC):=\min_{\calC\subset\mathbb{R}^d, |\calC|\le k}\sum_{i=1}^n\min_{c\in\calC}\dist(x_i,c)^z.\end{equation*}
In the streaming model, the points in $X$ arrive sequentially and a coreset $Z_t$ at time $t$ satisfies $\Cost(X_t,\calC)\in (1\pm \eps) \cdot \Cost(Z_t,\calC)$. 
We can then perform an efficient $(k,z)$-clustering algorithm on each coreset $Z_t$ to identify a near-optimal set of cluster centers, instead of operating on the full dataset arrived at time $t$. 
Naturally, smaller coresets correspond to more space-efficient algorithms. 
Another core usage of coresets is to $L_p$ subspace embeddings \cite{Woodruff14,CohenP15}, which compact a large-scale matrix $\bA$ while preserving all vector norms under projections, i.e., $\|\bM \bx\|_p \approx \|\bA \bx\|_p$ for all $\bx$, enabling efficient dimensionality reduction and accelerating down-stream tasks like least-squares regression. 

When constructed offline, many coreset schemes are known to require only $\poly_\calQ\left(d,\frac{1}{\eps} \right)$ samples \cite{Sarlos06,CohenP15,Cohen-AddadLSS22}.
As is often standard in the streaming literature, we allow each word of space to use $\Theta(\log(nd))$ bits of storage, so that we can store the weight and each coordinate of each point using $\O{1}$ words of space, and thus their space complexities depend only on structural parameters but not on $n$.
However, constructing and maintaining such coresets in the streaming model is far more challenging, since we do not have access to the entire dataset and require immediate decisions. 
Existing approaches, merge-and-reduce \cite{BentleyS80,Har-PeledM04} and online sampling \cite{Cohen-AddadWZ23,CohenMP20,WoodruffY23}, often incur logarithmic overhead in $n$ in space, which hinders their ability to handle large-scale datasets and puts forward the question of whether there is a separation in space between the offline and the streaming settings.
Thus it is natural to ask:
\begin{center}
Can we construct streaming algorithms for data summarization problems that achieve space usage (in words) independent of input size $n$?
\end{center}

\subsection{Our Contributions}
In this paper, we resolve the above question by providing a framework that produces space-efficient streaming algorithms via so called ``efficient encodings''.
Specifically, our framework efficiently encodes the coresets instead of storing them with high precision, reducing the memory consumption while enabling efficient reconstruction for future usage.
We then apply the framework to provide algorithms for $(k,z)$-clustering, $L_p$ subspace embeddings, and projection-cost preserving sketches with space independent of the input size $n$, without losing additional factors that depend on the structural properties, e.g., those related to $\calQ$, $d$, or $\eps$.
Therefore, our results match or improve on the space complexity of previous streaming algorithms in \emph{all} factors and match the \emph{offline} setting in the words of space. 
In essence, our framework demonstrates that streaming overhead is \emph{not} an inherent inevitability for data summarization problems, as long as they are amenable to an efficient encoding.

\paragraph{$(k,z)$-clustering.}
As mentioned earlier, $(k,z)$-clustering minimizes the clustering cost across all sets of at most $k$ centers, where the special cases $z=1$ and $z=2$ recover the classical $k$-median and $k$-means problems, respectively.
Intuitively, clustering partitions a dataset into groups so that points within the same group are similar, while points in different groups are dissimilar.  
Since its early study in the 1950s~\cite{steinhaus1956division,macqueen1967classification}, clustering has become a core unsupervised learning task in many domains, including bioinformatics, combinatorial optimization, computational geometry, computer graphics, data analysis, and machine learning.

In the (insertion-only) streaming model, the points $x_1,\ldots,x_n$ of $X$ arrive sequentially and for each time $t\in[n]$, the goal is to find a clustering $\calC_t$ on the prefix subset $X_t := \{x_1, \ldots, x_t\}$, inducing a cost that is within a $(1+\eps)$-fraction of the optimal clustering cost on $X_t$, for an accuracy parameter $\eps \in (0,1)$. 
We remark that due to the space constraints of the streaming model, we cannot represent input points in arbitrary precision and thus we generally assume the points in $X$ to lie within the grid $\{1,\ldots,\Delta\}^d$. 
Utilizing our efficient encoding framework, our first main result provides space-optimal coresets construction algorithms for $(k,z)$-clustering.
In the following, we use the notation $\tO{f}:=\polylog(f)$.
\begin{theorem}[Informal version of \thmref{thm:cluster:stream}]
\thmlab{thm:main}
Given a set $X$ of $n$ points on $[\Delta]^d$ and an accuracy parameter $\eps\in(0,1)$, there is a one-pass insertion-only streaming algorithm that uses $\tO{\frac{dk}{\min(\eps^4,\eps^{2+z})}}$ words of space and outputs a $(1+\eps)$-coreset of $X$. 
\end{theorem}
\thmref{thm:main} is the first result for $(k,z)$-clustering coreset construction on insertion-only streams using $\mathcal{O}_{k,d,\eps}(1)$ words of space, i.e., space usage independent of $n$. 
Despite a long line of work on clustering in the streaming model~\cite{Har-PeledM04,Har-PeledK07,Chen09,FeldmanL11,BravermanFLR19,stoc/BecchettiBC0S19,WoodruffZZ23}, the majority of previous results achieved space overheads that were polylogarithmic in $n$, compared to the offline setting. 
In a recent breakthrough work, \cite{Cohen-AddadWZ23} gave a streaming algorithm for $(k,z)$-clustering that uses $o_{k,d,\eps}(\log(n\Delta))$ words of space, but their result still fails to have full independence of $n$.  
On the contrary, \thmref{thm:main} matches the best known \emph{offline} coreset constructions~\cite{Cohen-AddadSS21,Cohen-AddadLSS22} in terms of dependencies on $d$, $k$, and $\frac{1}{\eps}$ in the space, and thus matches or improves upon the space usage of all previous streaming algorithms for insertion-only streams across all factors. 
Notably, our approach in \thmref{thm:main} supports an efficient encoding for any coreset construction. 
For instance, if $k<\frac{1}{\eps^2}$, it may be desirable to use existing coreset constructions of size $\tO{\frac{k^2}{\eps^2}}$ \cite{Cohen-AddadLSS22} rather than $\tO{\frac{k}{\eps^{z+2}}}$; our algorithm transitions to this case smoothly. 
In fact, our algorithm actually uses $\tO{dk\log(n\Delta)}+\frac{dk}{\min(\eps^4,\eps^{2+z})}\cdot\polylog\left(k,\frac{1}{\eps},\log(n\Delta)\right)$ bits of space, i.e., the $\frac{1}{\eps}$ factors do not multiply the $\log(n\Delta)$ factors. 
Therefore, our results are near-optimal with a recent lower bound on the communication complexity of clustering~\cite{zhu2024space}, which proved an $\Omega(dk/\varepsilon^2)$-bit lower bound for arbitrary clustering sketches under appropriate parameter regimes. 
See \figref{fig:space:summary} for a summary of the space complexity of recent results in the streaming model. 

\begin{figure}[!htb]
\centering
\small
{
\tabulinesep=1.1mm
\begin{tabu}{|c|c|}\hline
Streaming algorithm & Words of Memory \\\hline\hline
Online sensitivity sampling \cite{Cohen-AddadWZ23} & $\tO{\frac{d^2k^2}{\eps^2}\log^2 n}$ \\\hline
Merge-and-reduce with coreset of~\cite{Cohen-AddadLSS22} & $\tO{\frac{dk}{\min(\eps^4,\eps^{2+z})}\log^4 n}$ \\\hline
\cite{Cohen-AddadWZ23} & $\tO{\frac{dk}{\min(\eps^4,\eps^{2+z})}}\cdot\polylog(\log n)$ \\\hline\hline
\thmref{thm:main} (this work) & $\tO{\frac{dk}{\min(\eps^4,\eps^{2+z})}}$ \\\hline
\end{tabu}
}
\vspace{-0.05in}
\caption{Table of state-of-the-art $(k,z)$-clustering algorithms on insertion-only streams. We have $z=\O{1}$, $\Delta=\poly(n)$, and the assumption that $k>\frac{1}{\eps^z}$ for the purpose of presentation.}
\vspace{-0.1in}
\figlab{fig:space:summary}
\end{figure}

\paragraph{Subspace embeddings.}
Given a matrix $\bA\in\mathbb{R}^{n\times d}$ and an accuracy parameter $\eps\in(0,1)$, $L_p$ subspace embeddings seek for a matrix $\bM\in\mathbb{R}^{o(n)\times d}$ with fewer rows that preserves the projection of $\bA$ on every vector $\bx\in\mathbb{R}^d$ with respect to the $\ell_p$ norm, i.e., $\|\bM\bx\|_p \in (1\pm\eps)\|\bA\bx\|_p$.
Such an embedding reduces the dimension of $\bA$ while enabling nearly identical linear computations, including regression, low-rank approximation, and principal component analysis.  
Since their development in randomized linear algebra \cite{FriezeKV04,Woodruff14}, subspace embeddings have become a standard tool for large-scale data processing, supporting efficient algorithms in optimization, statistics, and machine learning.
We remark that the choice of the norm parameter $p$ determines which geometric structure is preserved.  
For instance, $p=2$ preserves Euclidean geometry, $p=1$ yields greater resilience to anomalous rows, and larger values of $p$ interpolate smoothly toward the behavior of $\ell_\infty$ norm.

In insertion-only streams, the rows of $\bA$ are given one by one in the stream.  
At each time step $t\in[n]$, the algorithm must maintain a correct $(1+\eps)$-embedding for the prefix matrix $\bA_t$ consisting of the first $t$ rows $\ba_1, \ldots, \ba_t$.
Our second result gives the construction of $\ell_p$ subspace embeddings:  
\begin{theorem}[Informal version of \thmref{thm:subspace:per:update}]
\thmlab{thm:lp:inf}
Given a matrix $\bA$ of $n$ rows on $[-M,\ldots,-1,0, 1,\ldots,M]^d$, with $M=\poly(n)$, there exists a one-pass streaming algorithm in the row arrival model that outputs a $(1+\eps)$-coreset. 
The algorithm uses $\tO{\frac{d^2}{\eps^2}}$ words of space for $p\in[1,2]$ and $\tO{\frac{d^{p/2+1}}{\eps^2}}$ words of space for $p>2$.
\end{theorem}
\vspace{-0.05in}
We note that existing algorithms based on using merge-and-reduce with the optimal coreset constructions for $L_p$ subspace embeddings require $\poly(d,\log n)$ bits of space, e.g., $\tO{\frac{d^2}{\eps^2}}\cdot\polylog(n)$ for $p\le 2$~\cite{DrineasMM06a,DrineasMM06b,Sarlos06} and $\tO{\frac{d^{p/2+1}}{\eps^2}}\cdot\polylog(n)$ for $p> 2$~\cite{CohenP15,WoodruffY23}.
Similarly, schemes based on online sampling also require $\poly(d,\log n)$ bits of space, e.g., $\tO{\frac{d^2}{\eps^2}}\cdot\polylog(n)$ for $p\le 2$~\cite{CohenMP20} and $\tO{\frac{d^{p/2+1}}{\eps^2}}\cdot\polylog(n)$ for $p>2$~\cite{CohenP15,BravermanDMMUWZ20,WoodruffY23}. 
Hence, it was unknown whether there is an inherent separation between the offline setting and the streaming model for coreset constructions for $L_p$ subspace embeddings. 
\thmref{thm:lp:inf} resolves this question, showing that there is no overhead for the streaming model, i.e., we can perform $L_p$ subspace embedding in $\mathcal{O}_{d,\eps}(1)$ words of space. 
We next remark on the differing behaviors across the ranges of $p$. 
This is not a coincidence, as \cite{LiWW21} showed that any constant-factor $L_p$ subspace embedding for $p\ge 2$ requires $\Omega(d^{\max(p/2,1)}+1)$ bits. 
Thus, our result has tight dependencies in terms of the dimension $d$. 
See \figref{fig:lp:space:summary} and \thmref{thm:subspace:per:update} for more details. 

\begin{figure}[!htb]
\centering
\small
{
\tabulinesep=1.1mm
\begin{tabu}{|c|c|}\hline
Streaming algorithm & Words of Memory \\\hline\hline
Merge-and-reduce with coreset of~\cite{DrineasMM06a,DrineasMM06b,Sarlos06}, $p\le 2$ & $\tO{\frac{d^2}{\eps^2}}\cdot\polylog(n)$ \\\hline
Online leverage score/Lewis weight sampling \cite{CohenMP20,WoodruffY23}, $p\le2$ & $\tO{\frac{d^2}{\eps^2}\log n}$ \\\hline
\thmref{thm:lp:inf} (this work), $p\le 2$ & $\tO{\frac{d^2}{\eps^2}}$ \\\hline
\hline
Merge-and-reduce with coreset of~\cite{DasguptaDHKM09}, $p\ge 1$ &$\tO{\frac{d^{\max(p/2+1,p)+2}}{\eps^2}\log^3 n}$ \\\hline
Merge-and-reduce with coreset of~\cite{CohenP15}, $p\ge 2$ &$\tO{\frac{d^{p/2+1}}{\poly(\eps)}}\cdot\polylog(n)$ \\\hline
Online Lewis weight sampling \cite{WoodruffY23}, $p>2$ & $\tO{\frac{d^{p/2+1}}{\eps^2}}\cdot\polylog(n)$ \\\hline
\thmref{thm:lp:inf} (this work), $p>2$ & $\tO{\frac{d^{p/2+1}}{\eps^2}}$ \\\hline
\end{tabu}
}
\vspace{-0.05in}
\caption{Table of $L_p$ subspace embedding algorithms on insertion-only streams. We summarize existing results with $\kappa=\poly(n)$ for the purpose of presentation.}
\figlab{fig:lp:space:summary}
\end{figure}
\paragraph{Projection-cost preserving sketch.}
Finally, we apply our efficient encoding framework to obtain projection-cost preserving sketches.
Given a matrix $\bA \in \mathbb{R}^{n \times d}$ and an accuracy parameter $\eps \in (0,1)$, the goal is to construct a matrix $\bB \in \mathbb{R}^{o(n) \times d}$ of fewer rows such that for all rank-$k$ orthogonal projection matrices $\bP \in \mathbb{R}^{d \times d}$,
$\|\bB-\bB \bP\|_F^2 \in(1\pm\eps)\|\bA-\bA \bP\|_F^2$,
where $\|\cdot\|_F$ denotes the Frobenius norm.
Introduced by \cite{CohenEMMP15}, projection-cost preserving sketches relax the requirements in subspace embeddings and instead preserve the residual cost under low-rank projections.
Importantly, these sketches can be used to compute the low-rank approximation of $\bA$, which is essential in random matrix algorithms and dimensionality reductions.
We next present our result:


\begin{theorem}[Informal version of \thmref{thm:pcp:per:update}]
\thmlab{thm:pcp}
Given a matrix $\bA$ of $n$ rows on $[-M,\ldots,-1,0,1,\ldots,M]^d$ with $M=\poly(n)$, there exists a one-pass streaming algorithm in the row arrival model that uses $\frac{k^2}{\eps^2} \cdot \polylog\left(\frac{1}{\eps},\log(nd\kappa)\right)+ \tO{\frac{dk}{\eps^2}}$ words of space and outputs a $(1+\eps)$ rank-$k$ projection-cost preservating sketch of $\bA$.
\end{theorem}
\vspace{-0.1in}

We note that \cite{BravermanDMMUWZ20} obtains a sketch with $\frac{dk}{\eps^2} \log n \log^2 \kappa$ words of space in the online setting, which naturally generalizes a streaming algorithm.
This complexity can be further reduced to $\frac{dk}{\eps^2}\polylog\left(\frac{1}{\eps},\log(nd\kappa)\right)$ words of space by passing to a merge-and-reduce structure with the optimal (offline) coreset construction in \cite{CohenMM17} (see e.g., the framework in \cite{Cohen-AddadWZ23}).
On the contrary, our streaming algorithm in \thmref{thm:pcp} achieves $\frac{k^2}{\eps^2} \cdot \polylog\left(\frac{1}{\eps},\log(nd\kappa)\right)+ \tO{\frac{dk}{\eps^2}}$ words of space.
By comparison, our result avoids $d$ multiplying the factors involving $n$, significantly improving the space complexity in previous algorithms.
Furthermore, assuming $d \gg k \poly(\log \log n)$, e.g., when the projection matrix has high dimension but low rank. the space complexity (in words) is entirely independent of $n$, matching the best known \emph{offline} coreset construction \cite{CohenMM17}.
In addition, for $\varepsilon=\Omega(k^{-1/2})$, PCP sketches formed from rescaled input rows require $\Omega(k/\varepsilon^2)$ rows in the worst case, as follows from the spectral-sparsification lower bound of~\cite{BatsonSS12}.
Storing such a sketch explicitly as a dense matrix uses $\Omega(dk/\varepsilon^2)$ entries.
When the $dk/\varepsilon^2$ term dominates our space bound, our space in words nearly-matches this offline lower bound.

\subsection{Technical Overview}
\applab{sec:overview}
In this section, we introduce our novel efficient encoding framework to achieve optimal space, and then we present the efficient encoding scheme for each problem.

\subsubsection{Streaming Framework for Optimal Space}
We begin with constructing the coreset in the offline setting, where the dataset $X$ is fixed and known by the algorithm.
For $X$ and a family of queries $\calQ$, the standard method is to sample each point $x \in X$ with probability proportional to its \emph{sensitivity}, i.e., $p(x) \propto s(x)$, and re-weight the point by the inverse $\frac{1}{p(x)}$ if sampled.
Here, the sensitivity of $x$ is defined by $s(x)=\max_{q\in \calQ}\frac{\Cost_q(x,C)}{\Cost_q(X,C)}$, where $\Cost_q(X,C)$ is the weighted sum of all elements in $X$; and captures the important elements in the dataset with respect to the queries $\calQ$, often leading to a good summarization of the entire dataset.

In the streaming setting, it is challenging to compute these sensitivities since we do not have access to the entire dataset.
The commonly-used online sensitivity approach samples each item $x_t$ of the stream with probability proportional to its \emph{online sensitivity}, which measures the importance of $x_t$ with respect to the points of the stream that have already arrived, i.e., its sensitivity with respect to $X_t=\{x_1,\ldots,x_t\}$. 
To compute the online sensitivities, we use the sensitivity of an element with respect to the previously sampled elements to give a close approximation, since we cannot afford to store the entire prefix $X_t$.
However, a crucial drawback of this approach is that future sampling probabilities now depend on whether the past elements are sampled, and so standard concentration bounds that require independence do not apply.
Although the correctness can be taken care of by martingale or decoupling analysis, these limitations lose poly-logarithmic factors in the number of samples, which is prohibitive for our purpose.

On the other hand, merge-and-reduce~\cite{BentleyS80,Har-PeledM04} constructs a binary tree on the data stream, where each node contains a block of points, and then ``merges'' two child nodes by offline coreset methods.
Formally, the approach framework first partitions the stream into consecutive blocks of size $f(\calQ,n,d,\eps',\delta')$, and then constructs a coreset with accuracy $(1+\eps')$ and failure probability $\delta'$ for each block of the stream. 
For each node in the tree at depth $\ell$, the framework takes the coresets of its two children at depth $\ell+1$ and applies the same offline summarization algorithm to their union, producing a new coreset with accuracy $(1+\eps')$. 
Observe that the coreset at the root of the tree represents a valid $(1+\eps)$-approximation for the entire stream for sufficiently small $\eps'$. 
However, the merge-and-reduce tree has height roughly $\log_{\calQ,d,\eps}(n)$ and the error in each level accumulates, so we need to construct offline coreset with finer accuracy, i.e., $\eps' = \frac{\eps}{\log_{\calQ,d,\eps}(n)}$, losing additional $\polylog_{\calQ,d,\eps}(n)$ factors, and thus, it cannot immediately be used to achieve our goal of streaming algorithms with space (in words) independent of $n$. 

Instead of directly implementing the merge-and-reduce method on the input stream, a recent framework proposed by \cite{Cohen-AddadWZ23} first down-sample the stream via the online sampling scheme and then passes the resulted stream $\calS'$ on to a merge-and-reduce method.
Now, the height of the merge-and-reduce coresets becomes $\O{\log(|\calS'|)} = \polylog_{\calQ,d,\eps}(\log(n))$, and so it suffices to have $\eps' = \frac{\eps}{\mathcal{O}(\log |\calS'|)} = \frac{\eps}{\polylog_{\calQ,d,\eps}(\log(n))}$, using less space compared to the previous choice of $\eps' = \frac{\eps}{\log_{\calQ,d,\eps}(n)}$. 
Although this method reduces $\polylog_{\calQ,d,\eps}(n)$ factors to $\polylog_{\calQ,d,\eps}(\log(n))$ factors, it still does not suffice for our goal of having space usage comparable to offline settings.

\paragraph{Efficient encoding framework.}
To solve the above problem, we propose a framework via efficient encoding schemes. 
Trivially, one can store the solution with high precision, but this would require $\O{\log(n)}$ bits for each item.
To achieve our goal of $\calO_{\calQ,d,\eps}(1)$ words of space, we design the efficient encoding scheme to store each item with less space.
Consider an ``active'' node in the merge-and-reduce tree (i.e., a node that is not merged to a parent node), our strategy is to store an inaccurate sketch, e.g., a constant-factor approximation coreset of the points in the node, with high precision.
Then we store the offset of the $(1+\eps)$-coreset obtained by the merge-and-reduce tree using quantization with precision parameter $\epsilon$, e.g., rounding the value to the nearest power of $(1+\epsilon)$.
Note that the size of the inaccurate sketch does not contain the additional $\polylog_{\calQ,d,\eps}(\log(n))$ factors, and so storing it with high precision does not incur factors involving $n$ in the number of words.
On the other hand, due to the efficient encoding scheme that stores the offsets, we avoid using $\log(n)$ bits to store each point.
Importantly, when we need to conduct some operations using the coresets, e.g., merging two coresets or computing the next online sensitivity, we can reconstruct the coresets from the efficient encoding effectively.
By appropriately choosing the precision parameter $\epsilon$, we ensure the accuracy of the reconstructed coresets while achieving $\calO_{\calQ,d,\eps}(1)$ bits of words.

Unfortunately, the merge-and-reduce tree implementing on $\calS'$ has height $\O{\log_{\calQ,d,\eps}(|S'|)}=\polylog_{\calQ,d,\eps}(\log(n))$ and therefore we need to efficiently encode $\polylog_{\calQ,d,\eps}(\log(n))$ active coresets simultaneously.
If we apply the efficient encoding scheme separately on each node, it requires storing the inaccurate sketch in each of them, and so the total space remains $\polylog_{\calQ,d,\eps}(\log(n))$, which does not meet our requirement.
To solve this, we implement a global efficient encoding scheme across all nodes.
Namely, we store a single inaccurate sketch of the entire stream with high precision, rather than storing an inaccurate sketch for each coreset, and then encode the offset of each simultaneously stored coreset to the sketch.
We observe that whenever we need to do a merge operation at some time $t$, we use the stored coreset to recompute a constant-factor approximation $C'$ to $X_t$, which is the entire prefix dataset arrived before $t$. 
For each coreset for a subset $S_v$ of data points representing an active node $v$ in the tree, we can then perform the efficient encoding with respect to $C'$ instead of a separate constant-factor approximation. 
Although each encoding no longer guarantees $(1+\eps)$-multiplicative approximation to $\Cost_q(S_v)$ for a query $q \in \calQ$, it still guarantees $\eps\cdot\Cost_q(X_t)$ additive error to $\Cost_q(S_v)$, which suffices for a $(1+\eps)$-approximation to $\Cost_q(X_t)$ after summing across all subsets $S_v$ at time $t$. 
In addition, if the recomputed sketch $C'$ suffices to approximate the online sensitivity of the next insertion, then the efficient encoding ensures the correctness of the entire framework.

We then show that our framework provides space-optimal streaming algorithms for various high-dimensional optimization problems; we believe that it might be applied to other problems, e.g., graph sparsification, and so it is of independent interest.
However, the efficient encoding schemes are not easy to obtain, especially, although the reconstructed sketch still gives a $(1+\eps)$-coreset, it might break the structural properties required to estimate the next online sampling probability.
For instance, being a $(1+\eps)$-projection-cost preservation sketch does not guarantee such properties directly, which needs further analysis.

\subsubsection{Efficient Encoding Schemes for Streaming Problems}
Despite our general framework, designing an efficient encoding remains highly nontrivial and fundamentally problem-dependent.

\paragraph{$(k,z)$-clustering.}
We first present the efficient encoding of the coresets for $(k,z)$-clustering. 
Suppose we have a set $C'$ of $\O{k}$ centers that is a constant-factor approximation to the optimal $(k,z)$-clustering on $X$. 
We can noisily encode $X$ by writing each $x\in X$ as $x=\pi_{C'}(x)+(x-\pi_{C'}(x))$, where $\pi_{C'}(x)$ is the closest center of $C'$ to $x$. 
We then round each coordinate of $x-\pi_{C'}(x)$ by rounding each coordinate to a power of $(1+\eps')$ for $\eps'=\poly\left(\eps,\frac{1}{d},\frac{1}{\log(n\Delta)}\right)$ to form a vector $y'$. 
Hence, given $C'$, each vector $x'=\pi_{C'}(x)+y'$ can be encoded using $\O{\log k+d\log\frac{1}{\eps'}}$ bits, since we can store the \emph{exponents} of the offsets. 
It is known~\cite{Cohen-AddadWZ23,zhu2024space} that if $X'$ is the set of all points of $X$ rounded in this manner, then $X'$ is a $(1+\eps)$-coreset for $X$. 
Moreover, we can round the weights of $X'$ to powers of $(1+\eps')$, thereby efficiently encoding each coreset in the merge-and-reduce framework to improve the overall space complexity of the problem. 
Recall that a local encoding scheme is insufficient to achieve optimal space to store the merge-and-reduce tree.
To overcome this issue and obtain a global efficient encoding, we provide a single constant-factor approximation to the global dataset, rather than providing a constant-factor approximation for each level of the merge-and-reduce tree.

\paragraph{Subspace embeddings.}
Next, we describe our approach for $L_p$ subspace embeddings.
It is not immediately obvious how to utilize a constant-factor subspace embedding $\bM\in\mathbb{R}^{m\times d}$ of the input matrix $\bA\in\mathbb{R}^{n\times d}$ to achieve an efficient encoding, especially since the dimensions are different. 
One possible approach is to project $\bA$ onto $\bM$ and round the resulting rows, but the rounding process could result in large error due to cancellation of rounded entries. 
For example, suppose that after the projection, there is a row $\br=(A, B)$ for some large values of $A,B>0$. 
By rounding each coordinate of $\br$ to the nearest power of $(1+\eps')$ for some $\eps'=\poly\left(\frac{\eps}{d,\log(nM)}\right)$, we obtain a row $\br'=(A',B')$. 
Now for $\bv=(B,-A)$, we have $\langle\bv,\br\rangle=0$, but $\langle\bv,\br'\rangle$ can be as large as $\eps'\cdot(|A|+|B|)$. 
Moreover, this error can compound across all rows of the matrix $\bA$ so that the resulting additive error to the estimate $\|\bA\bv\|_p^p$ could be as large as $(\eps')^p\cdot\|\bA\|_p^p$. 

We instead observe that the correct procedure is to first multiply the matrix $\bA$ with a preconditioner so that none of its rows can contribute a large amount to the error. 
To that end, we first use a well-conditioned basis to compute a preconditioner $\bP\in\mathbb{R}^{d\times d}$ of the constant-factor approximation $\bM$ and multiply each row of $\bA$, so that for all vectors $\bx\in\mathbb{R}^d$ with $\|\bA\bx\|_p=1$, we have that $\frac{1}{\poly(d)}\le\|\bA\bP\bx\|_p\le\poly(d)$. 
In particular, the maximum possible additive error $\langle\bv,\br'\rangle$ for each row $\br$ can be charged to the sensitivity of $\br$, and it follows that the sum of the errors due to the rounding can be at most $\eps'\cdot\poly(d)$. 
Since $\eps'=\poly\left(\frac{\eps}{d,\log(nM)}\right)$, then the overall error is $\O{\eps}$, which achieves a subspace embedding because $\|\bA\bx\|_p=1$. 
For a row $\bb_t=\ba_t\bP$, we then round each entry of $\bb_t$ to the nearest power of $(1+\eps')$ and store the exponent as before. 
We can then also achieve a streaming algorithm by compressing the entire merge-and-reduce tree, as before. 

\paragraph{Projection-cost preservation.}
To efficiently encode a matrix $\bA$ for the purpose of obtaining projection-cost preserving sketches, we consider its optimal rank-$k$ approximation $\bA_k$ and write the matrix as $\bA = \bA_k + (\bA - \bA_k)$.
We then store $\bA_k$ with high precision, e.g., $\O{\log n}$ bits per entry, and round each entry of the residual matrix $\bD = \bA - \bA_k$ to powers of $(1+\eps')$ to form a matrix $\bD'$.
We note that this is slightly different from our previous approach to round to a constant-factor approximation.
The intuition for rounding to $\bA_k$ is that projection-cost preservation depends only on how well the sketch represents distances in $k$-dimensional subspaces.
Decomposing $\bA = \bA_k + \bD$ isolates the structured component $\bA_k$ that captures the directions contributing most to the residual cost $\|\bA - \bA \bP\|_F^2$ for any rank-$k$ projection matrix $\bP$.
In contrast, the contribution of the residual matrix $\bD$ to the projection-cost is inherently bounded by $\|\bD\|_F^2 \le \|\bA - \bA \bP\|_F^2$ for any projection matrix $\bP$, and so quantizing $\bD$ with precision $\eps'$ gives us a desirable $(1+\eps)$-projection-cost preserving sketch in reconstruction.

We still need a global efficient encoding since we cannot afford to store the low-rank matrix $\bA_k$ for each node in the merge-and-reduce tree.
We observe that in the local efficient encoding, $\bA_k$ is a projection of the rows of $\bA$ onto the span of the top-$k$ right singular vectors.
Similarly, for global encoding, we project the PCP sketch matrix $\bQ_i$ on each node to the top-$k$ right singular vectors of $\bA$.
To obtain the efficient encoding, for each node $i$, we compute the head coefficients (corresponding to $\bA_k$) and store with high precision; and then we compute the tail residual (corresponding to $\bA - \bA_k$) and store with quantization (see the formal description in \appref{sec:pcp:all}).
As we mentioned earlier, being a projection-cost preserving sketch does not necessarily ensure that it can be used to provide a valid estimation of the online sampling probability, as the latter would require subspace embedding properties.
Fortunately, we show that by properly choosing the quantization factor, we are able to satisfy this requirement, giving the correctness of the overall algorithm.

\subsection{Preliminaries}

For a positive integer $n$, we write $[n]=\{1,2,\ldots,n\}$.  
We use $\poly(n)$ to denote an arbitrary fixed polynomial in $n$, and $\polylog(n)$ to denote a polynomial in $\log n$.  
We use the notation $\tO{f}:=\polylog(f)$.
An event is said to hold with high probability if it occurs with probability at least $1-\frac{1}{\poly(n)}$.

For $(k,z)$-clustering, we assume the dataset $X \subset [\Delta]^d$.
Following standard assumptions in the streaming model, we assume $\Delta=\poly(n)$ and allow each memory word to store $\Theta(\log(nd\Delta))$ bits, i.e., $\Theta(\log(ndM))$ bits for for subspace embedding and projection-cost preservation.  
This ensures that the weight and each coordinate of a point in $\mathbb{R}^d$ can be represented using $\O{1}$ words.

Throughout the paper, we restrict our attention to Euclidean spaces rather than general metric spaces.  
For points $x,y\in\mathbb{R}^d$, we write $\dist(x,y)=\|x-y\|_2$.
For a point $x$ and a subset $S\subset\mathbb{R}^d$, we further define $\dist(x,S)=\min_{s\in S}\dist(x,s)$.  
For $z\ge 1$, the $\ell_z$ norm of a vector $x$ is denoted by $\|x\|_z$, where $\|x\|_z^z=\sum_{i=1}^d |x_i|^z$.  
Given a matrix $\bA\in\mathbb{R}^{n\times d}$, its Frobenius norm is
\[
\|\bA\|_F^2=\sum_{i\in[n]}\sum_{j\in[d]} A_{i,j}^2.
\]
For a fixed $z\ge 1$ and sets $X,C\subset\mathbb{R}^d$ with $X=\{x_1,\ldots,x_n\}$, we define the clustering objective
\[
\Cost(X,C)=\sum_{i=1}^n \dist(x_i,C)^z.
\]
We will make use of the following inequality, sometimes referred to as a generalized triangle inequality.

\begin{fact}[Generalized triangle inequality]
\factlab{fact:triangle}
For any $z\ge 1$ and any $x,y,w\in\mathbb{R}^d$,
\[
\dist(x,y)^z \le 2^{z-1}\big(\dist(x,w)^z+\dist(w,y)^z\big).
\]
\end{fact}

We also utilize the classical Johnson-Lindenstrauss dimensionality reduction result.

\begin{theorem}[Johnson-Lindenstrauss lemma]
\cite{johnson1984extensions}
\thmlab{thm:jl}
Given an accuracy parameter $\eps\in\left(0,\tfrac12\right)$ and set $m=\O{\frac{1}{\eps^2}\log n}$.  
For any set $X\subset\mathbb{R}^d$ of $n$ points, there exists a distribution over linear maps $\Pi:\mathbb{R}^d\to\mathbb{R}^m$ such that, with high probability over $\pi\sim\Pi$,
\[
(1-\eps)\|x-y\|_2 \le \|\pi x-\pi y\|_2 \le (1+\eps)\|x-y\|_2
\]
holds simultaneously for all $x,y\in X$.
\end{theorem}

We further recall the standard Hoeffding's inequality.

\begin{theorem}[Hoeffding's inequality]
\thmlab{thm:hoeffding's:inequality}
Let $X_1,\ldots,X_n$ be independent random variables with $a_i\le X_i\le b_i$, and let $S_n=\sum_{i=1}^n X_i$.  
Then for any $t>0$,
\[
\PPr{|S_n-\Ex{S_n}|>t}
\le
2\exp\!\left(-\frac{t^2}{\sum_{i=1}^n (b_i-a_i)^2}\right).
\]
\end{theorem}

Next we formalize the notion of a coreset, which serves as a compact weighted proxy for a dataset while approximately preserving a class of query costs.

\begin{definition}[Coreset]
\deflab{def:coreset:frame}
Let $\calQ$ be a family of query functions over a weighted point set $X\subseteq[\Delta]^d$ with weights $w:X\to\mathbb{R}_{>0}$.  
A weighted subset $Y\subseteq X$ is a $(1+\varepsilon)$-coreset for $(X,\calQ)$ if, for every $q\in\calQ$,
\[
(1-\varepsilon)\Cost(q,X)
\le
\Cost_q(Y)
\le
(1+\varepsilon)\Cost(q,X),
\]
where $\Cost_q(Y)=\sum_{x\in Y} w(x)\,q(x)$.
\end{definition}

This definition subsumes the specific coresets considered in this work.  
For $(k,z)$-clustering, a query $q$ corresponds to a set of $k$ centers, and $q(x)$ equals the $z$-th power of the distance from $x$ to its closest center.  
For subspace embeddings, $q$ is a vector $\bv\in\mathbb{R}^d$ and $q(x)=|\bx^\top \bv|^p$, so that $\Cost_q(Y)=\|\bY\bv\|_p^p$ for the matrix $\bY$ formed by rows of $Y$.  
For projection-cost preservation (PCP), $q$ is a projection matrix $\bP$, and the cost is the residual Frobenius norm $\|\bY-\bY\bP\|_F^2$.

\section{Frameworks for Space-Optimal Streaming Algorithms}
\seclab{sec:frame:space:front}

In this section, we formally describe our efficient encoding framework.
Suppose that we have the following subroutines, each with failure probability at most $\delta$:
\begin{itemize}
    \item An online sampling scheme that returns a $(1+\eps)$-coreset of size $\calS_{\text{ol}}(\eps,\delta)$.
    \item An algorithm that merges two coresets to obtain a $(1+\eps)$-coreset of size $\calS_{\text{off}}(\eps,\delta)$.
    \item A global efficient encoding scheme that encodes a stream with $m$ partitions $\calK = \calK_1 \circ \ldots \circ \calK_m$, where each part $\calK_i$ has size at most $\calS_{\text{off}}(\eps,\delta)$, using $\calS_{\text{gl-encode}}(\eps,\delta) + m \cdot \calS_{\text{loc-encode}}(\eps,\delta)$ bits of memory, where $\calS_{\text{gl-encode}}(\eps,\delta)$ denotes the space to store the global constant-factor approximation and $\calS_{\text{loc-encode}}(\eps,\delta)$ denotes the space to store the offset of each active coreset.
\end{itemize}

Note that we do not show the dependency of $n$ and $d$ in the above space complexities for simplicity of presentation.
We then assemble the above subroutines.
Given a stream $\calS$ of length $n$, we use the online sampling scheme to generate a stream $\calS'$ and pass it on to a merge-and-reduce tree to produce a coreset $\calC$.
We store the active nodes in the tree using global efficient encoding.
Upon each new arrival, we reconstruct the coreset from the global encoding, which defines the online sampling probability.
We visualize this process in \figref{fig:outline}.

\begin{figure*}[!htb]
\begin{mdframed}
\begin{enumerate}
\item
Given a stream $\calS$ of length $n$, use an online sampling scheme to generate a stream $\calS'$. 
\item
Run merge-and-reduce on $\calS'$ to produce a coreset $\calC$. 
\item
Given a constant-factor approximation $\calC'$ of $\calS$, round the active nodes in the merge-and-reduce data structure to $\calC'$ with global efficient encoding.
\item 
Whenever a new item arrives in the stream, reconstruct the coreset from the efficient encoding to compute the online sampling probability in Step (1).
\end{enumerate}
\end{mdframed}
\vspace{-0.1in}
\caption{High-level summary of our approach}
\figlab{fig:outline}
\end{figure*}


Note that there are at most $\ell= \frac{\calS_{\text{ol}}(\eps, \delta)}{\calS_{\text{off}}(\eps, \delta)}$ leaf nodes in the merge-and-reduce tree, so the number of active nodes is bounded by the height $h = \O{\log \ell}$.
Since each level in the merge-and-reduce tree at most incurs an additional $(1+\eps)$-factor in the multiplicative error, it suffices to take $\eps' = \O{\frac{\eps}{h}}$ and $\delta' = \O{\frac{\delta}{\ell}}$ when we construct each coreset.
Therefore, the global encoding scheme encodes $h$ coresets and uses $\calS_{\text{gl-encode}}\left(\frac{\eps}{h},\frac{\delta}{\ell}\right) + h \cdot \calS_{\text{loc-encode}}\left(\frac{\eps}{h},\frac{\delta}{\ell}\right)$ bits of space in total.

We next formally present and analyze the above framework.
To begin with, we state the formal definitions of the local and global efficient encoding schemes.

\begin{definition}[Efficient encoding scheme]
Given a dataset $X$ and a family of query functions $\mathcal{Q}$, an efficient encoding scheme stores a space-efficient structured representation $C(X)$ such that one can recover a $(1+\eps)$-coreset $X'$ of $X$ from $C(X)$.
\end{definition}


\begin{definition}[Global efficient encoding]
Given a stream $\mathcal{S} = \mathcal{S}_1 \circ \mathcal{S}_2 \circ, \ldots , \circ \calS_m$, a global efficient encoding scheme stores a sketch $C(\calS_i)$ for each $\calS_i$ such that one can recover a $(1+\eps)$-coreset $\calS'$ of $\calS$ from $\cup_i C(\calS_i)$.
\end{definition}

As  we discussed earlier, storing each active coreset separately with the local efficient encoding scheme loses extra $\polylog_{\calQ,d,\eps}(\log(n))$ factors, so the global efficient encoding scheme is necessary to avoid this depencence.
The global encoding no longer gives a $(1+\eps)$-coreset for each node in the tree by reconstruction, however, it gives a $(1+\eps)$-coreset for the entire stream, which is sufficient for our purpose.
We assembly all the subroutines in \algref{alg:framework:space}.

\begin{algorithm}[!htb]
\caption{Space-Optimal Streaming Framework}
\alglab{alg:framework:space}
\begin{algorithmic}[1]
\Require{Stream $\mathcal{S}$ of $n$ objects in $\mathbb{R}^d$ (points, rows, or vectors), target query function family $\calQ$ defining $\Cost_{\calQ}(\cdot)$, accuracy parameter $\eps\in(0,1)$, failure probability $\delta$, budget $B$ for the merge-and-reduce coreset}
\Ensure{$(1+\eps)$-coreset of $\mathcal{S}$}
\For{each element $x_t\in\mathcal{S}$}
    \State Reconstruct the current coreset from the global encoding
    \State Sample $x_t$ with the online sampling scheme and pass on a sub-stream $\calS'$. \Comment{Compute the sampling probability using the reconstructed coreset}
    \If{$x_t$ is sampled to $\calS'$}
    \State{Input $x_t$ to the on-going leaf node in the merge-and-reduce tree}
    \State{Store $x_t$ with high precision, (e.g., $\O{\log n}$ bits for each entry)}
    \If{the on-going leaf node has size $B$}
    \State{Let $\calC_1, \ldots, \calC_\ell$ be the active coresets. Reconstruct $\calC_1, \ldots, \calC_\ell$ from the global encoding} \Comment{$\calC_1, \ldots, \calC_\ell$ have not been merged to a parent node. $\bigcup_{i \in \ell} \calC_i$ is a coreset of the previous stream}
    \State{Let $\calC_{\ell+1}$ be the coreset on the on-going leaf node, merge $\calC_{\ell+1}$ with $\calC_1, \ldots, \calC_\ell$} \Comment{Merge two child nodes using the offline algorithm}
    \State{Store the resulted coresets with global efficient encoding}
    \EndIf
    \EndIf
\EndFor
\State \Return{the items in the active coresets}
\end{algorithmic}
\end{algorithm}

We provide guarantees of \algref{alg:framework:space} in \frameref{frame:space}.
\begin{framework}
\framelab{frame:space}
Given a stream $\calS$ of length $n$, a family of query functions $\calQ$, and an accuracy parameter $\eps$, suppose that we have access to the following subroutines:
\begin{itemize}
    \item An online sampling scheme that returns a $(1+\eps)$-coreset of size $\calS_{\text{ol}}(\calQ,n,d,\eps,\delta)$ with probability $1-\delta$.
    \item An (offline) algorithm that merges two coresets of size $\calS_{\text{off}}(\calQ,n,d,\eps,\delta)$ to obtain a $(1+\eps)$-coreset of size $\calS_{\text{off}}(\calQ,n,d,\eps,\delta)$ with probability $1-\delta$.
    \item A global efficient encoding scheme that encodes a stream with $m$ partitions $\calK = \calK_1 \circ \ldots \circ \calK_m$, where each part $\calK_i$ has size at most $\calS_{\text{off}}(\calQ,n,d,\eps,\delta)$, using $\calS_{\text{gl-encode}}(\calQ,n,d,\eps,\delta) + m \cdot \calS_{\text{loc-encode}}(\calQ,n,d,\eps,\delta)$ bits of memory, where $\calS_{\text{gl-encode}}(\calQ,n,d,\eps,\delta)$ denotes the space to store the global constant-factor approximation and $\calS_{\text{loc-encode}}(\calQ,n,d,\eps,\delta)$ denotes the space to store the offset of each active coreset.
\end{itemize}
Then there is an algorithm that returns a $(1+\eps)$-coreset at all times with probability $1- \delta$.
The algorithm uses 
\[\calS_{\text{gl-encode}}\left(\calQ,n,d,\frac{\eps}{h},\frac{\delta}{\ell}\right) + h \cdot \calS_{\text{loc-encode}}\left(\calQ,n,d,\frac{\eps}{h},\frac{\delta}{\ell}\right)\]
bits of memory, where $\ell = \frac{\calS_{\text{ol}}(\calQ,n,d,\eps,\delta)}{\calS_{\text{off}}(\calQ,n,d,\eps,\delta)}$ is the number of leaf nodes in the merge-and-reduce tree and $h = \O{\log \ell}$ is the height.
Here, we assume that the space complexities scale at most polynomially with $\eps$ and $\delta$, and so we omit the constant terms.
\end{framework}
\begin{proof}
We analyze the correctness and the complexity of \algref{alg:framework:space}.
Let $B = \calS_{\text{off}}(\calQ,n,d,\eps,\delta)$ be the budget for the merge-and-reduce coreset.
First, we apply the online sampling scheme which succeeds upon each arrival with probability $1- \frac{\delta}{\poly(n)}$ so that we can union bound across the entire stream.
Note that we reconstruct the coreset from the global encoding scheme at each step, which ensures the validity of the online sampling probability.

Then we implement the merge-and-reduce method on the online sampled sub-stream with coreset size $\calS_{\text{off}}(\calQ,n,d,\eps,\delta)$.
Since the online sampling scheme returns a stream of size $\calS_{\text{ol}}(\calQ,n,d,\eps,\delta)$ (which we do not store), the number of leaf nodes is at most $\ell = \frac{\calS_{\text{ol}}(\calQ,n,d,\eps,\delta)}{\calS_{\text{off}}(\calQ,n,d,\eps,\delta)}$ and the height of the merge-and-reduce tree is $h = \O{\log \ell}$.
Since our definition of the coreset attains mergeability, i.e., if $A_1$ is a $(1+\eps)$-coreset of $B_1$ and $A_2$ is a $(1+\eps)$-coreset of $B_2$, then $A_1 \circ A_2$ is a $(1+\eps)$-coreset of $B_1 \circ B_2$, and transitivity, i.e., if $A$ is a $(1+\eps)$-coreset of $B$ and $B$ is a $(1+\eps)$-coreset of $C$, then $A$ is a $(1+\eps)^2$-coreset of $C$ (see \defref{def:coreset:frame}).
Thus, it suffices to apply the offline algorithm to merge the coresets with parameters $\eps' = \theta\left(\frac{\eps}{h}\right)$ and $\delta' = \theta\left(\frac{\delta}{\ell}\right)$.
Then the total distortion is at most $(1+\eps')^h = 1+\O{\eps' h} = 1+\eps$, and the failure probability can be union bounded since there are at most $\O{\ell}$ times of merging coresets.
This gives the correctness of our algorithm.

Next, we compute the bits of memory storage.
Note that when we apply the global efficient encoding scheme, there are at most $h$ coresets to encode (which is the number of coresets that are not merged to a parent node in the tree) and each of size $\calS_{\text{off}}\left(\calQ,n,d,\frac{\eps}{h},\frac{\delta}{\ell}\right)$.
Therefore, the total number of bits we store is
\[\calS_{\text{gl-encode}}\left(\calQ,n,d,\frac{\eps}{h},\frac{\delta}{\ell}\right) + h \cdot \calS_{\text{loc-encode}}\left(\calQ,n,d,\frac{\eps}{h},\frac{\delta}{\ell}\right).\]
\end{proof}

We remark that in the high-dimensional optimization problems we study, the state-of-the-art online algorithms and their offline counterparts are often off by $\polylog_{\calQ,d,\eps}(n)$, and so in the above formula, $h = \log \frac{\calS_{\text{ol}}(\calQ,n,d,\eps,\delta)}{\calS_{\text{off}}(\calQ,n,d,\eps,\delta)} = \polylog_{\calQ,d,\eps}(\log(n))$.
Recall that $\calS_{\text{gl-encode}}\left(\calQ,n,d,\frac{\eps}{h},\frac{\delta}{\ell}\right)$ is the space to store the global constant-factor approximation, which is often tight in words of space compared to the offline setting.
Moreover, $\calS_{\text{loc-encode}}\left(\calQ,n,d,\frac{\eps}{h},\frac{\delta}{\ell}\right)$ is the space to store the offset of each active coreset, which only loses $\polylog_{\calQ,d,\eps}(\log(n))$ factors by our efficient quantization methods.
Therefore, the second term in our space complexity only loses $\polylog_{\calQ,d,\eps}(\log(n))$ factors, which becomes a lower-order term, i.e., $o(1)$, in words of space as we assume each word stores $\log(n)$ bits.
Thus, our framework gives $\calO_{\calQ,d,\eps}(1)$ words of space as we desire.

\section{\texorpdfstring{$(k,z)$}{(k,z)}-Clustering}
\applab{app:k:z}
In this section, we apply our framework in \secref{sec:frame:space:front} to achieve optimal $\mathcal{O}_{k,d,\eps}(1)$ words of space. 
Particularly, we provide the global efficient encoding for coresets for $(k,z)$-clustering and show its usage in the framework.

We begin by reviewing the formal definition of a strong coreset for $(k,z)$-clustering.  
\begin{definition}[Coreset]
\deflab{def:coreset}
Given a set $X = \{x_1,\ldots,x_n\}$ of points in $\mathbb{R}^d$, a distance function $\dist$, and an accuracy parameter $\eps>0$, a \emph{coreset} for $(k,z)$ clustering is a weighted subset $S$ such that for any set $C \subset \mathbb{R}^d$ of $k$ points, we have
\[(1-\eps)\sum_{t=1}^n\dist(x_t,C)^z\le\sum_{q\in S}w(q)\dist(q,S)^z\le(1+\eps)\sum_{t=1}^n\dist(x_t,C)^z,\]
where $w$ is the weight function.
\end{definition}

We next define the (offline) sensitivity for $(k,z)$-clustering. 
\begin{definition}[Sensitivities for $(k,z)$-clustering]
\deflab{def:sens:cluster}
The \emph{sensitivity} of a point $x\in X$ for $(k,z)$-clustering in a metric space with metric $\dist$ is 
\[\max_{C:|C|\le k}\frac{\Cost(x,C)}{\Cost(X,C)}=\max_{C:|C|\le k}\frac{\dist(x,C)^z}{\sum_{x\in X}\dist(x,C)^z}.\]
\end{definition}

We use the following (offline) algorithm for $(k,z)$-clustering:
\begin{theorem}\cite{Cohen-AddadLSS22,huang2023optimal,nips/Cohen-AddadLSSS22}
\thmlab{thm:offline:coreset:size}
Given an accuracy parameter $\eps\in(0,1)$, there exists an algorithm that samples $\tO{\min\left(\frac{1}{\eps^2}\cdot k^{2-\frac{z}{z+2}},\frac{1}{\min(\eps^4,\eps^{2+z})}\cdot k\right)}$ points and constructs a $(1+\eps)$-coreset. 
\end{theorem}

We next define the online sensitivity for $(k,z)$-clustering. 
\begin{definition}[Online sensitivity for $(k,z)$-clustering]
\deflab{def:online:sens:cluster}
Given a sequence of points $x_1,\ldots,x_n$, the \emph{online sensitivity} of $x_t$ for $(k,z)$-clustering is
\[\max_{C:|C|\le k}\frac{\Cost(x_t,C)}{\Cost(X_t,C)}=\max_{C:|C|\le k}\frac{\dist(x_t,C)^z}{\sum_{i=1}^t\dist(x_i,C)^z},\]
where $X_t:=\{x_1,\ldots,x_t\}$ is the prefix dataset at time $t$.
\end{definition}
We next recall the following upper bound on the sum of the online sensitivities, which is used to bound the number of samples. 
\begin{theorem}
\thmlab{thm:total:online:sens}
\cite{Cohen-AddadWZ23}
Given a sequence of points $x_1,\ldots,x_n \subset [\Delta]^d$ and let $\sigma(x_t)$ denote the online sensitivity of $x_t$ for each $t\in[n]$ for $(k,z)$-clustering, where $z\ge 1$, then we have
\[\sum_{t=1}^n\sigma(x_t)=\O{2^{2z}k\log^2(nd\Delta)}.\]
\end{theorem}
We next state the formal guarantees of online sensitivity sampling.
\begin{theorem}[Online sensitivity sampling]
\thmlab{thm:online:sens:cluster}
\cite{Cohen-AddadWZ23}
Given a sequence of points $x_1,\ldots,x_n \subset [\Delta]^d$, we sample each point $x_t$ with probability $p_t\ge\min(1,\gamma\cdot\sigma(x_t))$, where $\sigma_t$ is the online sensitivity of $x_t$ and $\gamma=\O{\frac{dk}{\eps^2}\log\frac{n\Delta}{\eps}}$ and weight $x_t$ by $\frac{1}{p_t}$ if sampled. 
Then with high probability, we obtain a weighted subset that is a $(1+\eps)$-coreset for $(k,z)$-clustering containing at most $\O{\frac{dk^2}{\eps^2}\log^3\frac{n\Delta}{\eps}}$ points. 
\end{theorem}

\subsection{Efficient Encoding for Coreset Construction for \texorpdfstring{$(k,z)$}{(k,z)}-Clustering}
We now give our efficient encoding for a given coreset for $(k,z)$-clustering. 
Given a dataset $X$, which can be viewed as either the original input or a set of weighted points that forms a coreset of some underlying dataset, we first acquire a constant-factor approximation $C'$ for $(k,z)$-clustering on $X$. 
For each $x\in X$, let $\pi_{C'}(X)$ be the closest center of $C'$ to $x$. 
We then write each $x\in X$ as $x=\pi_{C'}(x)+(x-\pi_{C'}(x))$, thereby decomposing $x$ into the closest center $\pi_{C'}(x)$ and its offset $x-\pi_{C'}(x)$ from the center. 
We show that for the purposes of $(1+\eps)$-approximation for $(k,z)$-clustering, we can afford to round each coordinate of $x-\pi_{C'}(x)$ to a power of $(1+\eps')$, where $\eps'=\poly\left(\eps,\frac{1}{d},\frac{1}{\log(n\Delta)}\right)$, forming a vector $y'$. 
Finally, we can store $C'$ and thus encode the vector $x'=\pi_{C'}(x)+y'$ using $\O{\log k+d\log\left(\frac{1}{\eps},d,\log(n\Delta)\right)}$ bits, by storing the identity of $\pi_{C'}(x)$ and the exponent of the offset for each of the $d$ coordinates. 
We give the algorithm in full in \algref{alg:coreset:cluster}. 

\begin{algorithm}[!htb]
\caption{Efficient Encoding for Coreset Construction for $(k,z)$-Clustering}
\alglab{alg:coreset:cluster}
\begin{algorithmic}[1]
\Require{Data set $X\subset[\Delta]^d$ with weight $w(\cdot)$, accuracy parameter $\eps\in(0,1)$, number of clusters $k$, parameter $z\ge 1$, failure probability $\delta\in(0,1)$}
\Ensure{$(1+\eps)$-coreset for $(k,z)$-clustering}
\State{$\eps'\gets\frac{\poly(\eps^z)}{\poly(k,\log(nd\Delta)}$}
\State{Find a set $C'$ of $k$ centers that is a constant-factor approximation to $(k,z)$-clustering on $X$}
\For{each $x\in X$}
\State{Let $c'(x)$ be the closest center of $C'$ to $x$}
\State{Let $y'$ be the offset $x-c'(x)$}
\State{Let $y$ be $y'$ with coordinates rounded to a power of $(1+\eps')$}
\State{Let $x'=(c'(x),y)$, storing the \emph{exponent} for each coordinate of $y$}
\State{$X'\gets X'\cup\{x'\}$}
\EndFor
\State{\Return $(C',X')$}
\end{algorithmic}
\end{algorithm}

To show correctness of our encoding, we first recall the following fact.
\begin{fact}[Claim 5 in \cite{SohlerW18}]
\factlab{fact:gen:tri}
Suppose $z\ge 1$, $x,y\ge 0$, and $\eps\in(0,1]$.  
Then
\[(x+y)^z\le(1+\eps)\cdot x^z+\left(1+\frac{2z}{\eps}\right)^z\cdot y^z.\]
\end{fact}

Given a constant-factor approximation $C'$, we now show that the rounding $X'$ of the dataset $X$ by representing each point as a rounded offset from its closest center in $C'$ is a strong coreset of $X$. 
\begin{lemma}
\lemlab{lem:x:to:xprime:cluster}
Let $\eps\in\left(0,\frac{1}{2}\right)$ and let $X'$ be the weighted dataset $S$ defined by their offsets from the set $C'$ of centers from \algref{alg:coreset:cluster}. 
Then for all $C\subset[\Delta]^d$ with $|C|\le k$, 
\[(1-\eps)\cdot\Cost(C,X)\le\Cost(C,X')\le(1+\eps)\cdot\Cost(C,X).\]
\end{lemma}
\begin{proof}
Let $C$ be any fixed set of at most $k$ centers. 
We abuse notation so that for each $x\in X$, we use $x'$ to denote the corresponding point $c(x)+y$, where $y'=x-c'(x)$ is the offset, and $y$ is $y'$ with its coordinates rounded to a power of $(1+\eps')$. 
Similarly, we use $c'$ to denote $c(x)\in C'$. 
Thus, we have $\|x-x'\|_2\le\eps'\cdot\|c'-x\|_2$ for all $x\in X$. 

For ease of presentation, we first write $X'$ so that the weights are not rounded to a power of $(1+\eps')$. 
By the triangle inequality,
\[\Cost(C,X')=\sum_{x'\in X'}(\dist(x',C))^z\le\sum_{x'\in X'}(\dist(x',x)+\dist(x,C))^z.\]
Since $\dist(x,x')=\|x-x'\|_2\le\eps'\cdot\|c'-x\|_2=\eps'\cdot\dist(x,C')$, then
\begin{align*}
\Cost(C,X')&\le\sum_{x'\in X'}\left(\eps'\cdot\dist(x,C')+\dist(x,C)\right)^z.
\end{align*}
By \factref{fact:gen:tri},
\begin{align*}
\Cost(C,X')&\le\sum_{x'\in X'}\left(\left(1+\frac{\eps}{2}\right)\cdot(\dist(x,C))^z+\left(1+\frac{4z}{\eps}\right)^z\cdot(\eps'\cdot\dist(x,C'))^z\right).
\end{align*}
Thus,
\[\Cost(C,X')\le\left(1+\frac{\eps}{2}\right)\cdot\Cost(C,X)+\left(1+\frac{4z}{\eps}\right)^z(\eps')^z\cdot\Cost(C',X).\]
Since $C'$ is a constant-factor approximation to the optimal $(k,z)$-clustering of $X$, then there exists a constant $\gamma\ge 1$ such that $\Cost(C,X')\le\gamma\cdot\Cost(C,X)$. 
Hence, 
\[\Cost(C,X')\le\left(\left(1+\frac{\eps}{2}\right)+\left(1+\frac{4z}{\eps}\right)^z(\eps')^z\cdot\gamma\right)\cdot\Cost(C,X).\]
Then for $\eps'=\frac{\poly(\eps^z)}{\poly(k,\log(nd\Delta)}$, we have that
\[\Cost(C,X')\le\left(1+\frac{2}{3}\eps\right)\cdot\Cost(C,X),\]
which almost gives the right hand side of the inequality (due to the notation $X'$ representing the output prior to the rounding of the weights). 

Similarly, by triangle inequality and $\dist(x,x')\le\eps'\cdot\dist(x,C')$,
\begin{align*}
\Cost(C,X)&=\sum_{x\in X}(\dist(x,C))^z\\
&\le\sum_{x\in X}(\dist(x,x')+\dist(x',C))^z\\
&\le\sum_{x'\in X'}\left(\eps'\cdot\dist(x,C')+\dist(x',C)\right)^z.
\end{align*}
By \factref{fact:gen:tri},
\begin{align*}
\Cost(C,X)&\le\left(1+\frac{\eps}{2}\right)\cdot\Cost(C,X')+\left(1+\frac{4z}{\eps}\right)^z(\eps')^z\cdot\Cost(C',X)\\
&\le\left(1+\frac{\eps}{2}\right)\cdot\Cost(C,X')+\left(1+\frac{4z}{\eps}\right)^z(\eps')^z\cdot\gamma\cdot\Cost(C,X).
\end{align*}
Then for $\eps'=\frac{\poly(\eps^z)}{\poly(k,\log(nd\Delta)}$, we have that
\[\Cost(C,X)\le\left(1+\frac{\eps}{2}\right)\cdot\Cost(C,X')+\frac{\eps}{100}\cdot\Cost(C,X),\]
so that $\left(1-\frac{2}{3}\cdot\eps\right)\Cost(C,X)\le\Cost(C,X')$, which almost gives the left-hand side of the inequality (due to the notation $X'$ representing the output prior to the rounding of the weights). 

Finally, since the weights are rounded to a power of $(1+\eps')$, then the cost can only change by $(1+\eps')$. 
Since $\eps'=\frac{\poly(\eps^z)}{\poly(k,\log(nd\Delta)}$, then it follows that
\[(1-\eps)\cdot\Cost(C,X)\le\Cost(C,X')\le(1+\eps)\cdot\Cost(C,X).\]
\end{proof}
\noindent
We now give the full guarantees of \algref{alg:coreset:cluster}, which gives an efficient encoding of an input set $X$.  
\begin{lemma}
\lemlab{lem:efficient:coreset:cluster}
Let $X$ be a coreset construction with weights bounded by $[1,\poly(nd\Delta)]$. 
Then $X'$ is a $(1+\eps)$-coreset for $X$ that uses 
$\O{dk\log(n\Delta)}+|X|\cdot\polylog\left(k,\frac{1}{\eps},\log(nd\Delta),\log\frac{1}{\delta}\right)$ bits of space. 
\end{lemma}
\begin{proof}
\algref{alg:coreset:cluster} then outputs a set $(C',X')$ that encodes $X'$ using $C'$. 
By \lemref{lem:x:to:xprime:cluster}, we have that for all $C\subset[\Delta]^d$ with $|C|\le k$, 
\[(1-\eps)\cdot\Cost(C,X)\le\Cost(C,X')\le(1+\eps)\cdot\Cost(C,X).\]
Thus, $X'$ is a strong coreset for $X$. 
 
It remains to analyze the space complexity of $(C',X')$. 
We first require $C'$ to encode $S$. 
Since $C'$ is a set of $k$ centers, then $C'$ can be represented using $\O{dk\log(n\Delta)}$ bits of space. 
Additionally, each point $x'$ in $X'$ is encoded using $\O{d\log\left(\frac{1}{\eps}+\log n+\log\Delta\right)}$ bits of space due to storing the $d$ coordinates of the rounded offset $y$. 
In particular, we do not store the explicit coordinates, but rather their exponents. 
Moreover, we use $\log k$ bits to store the closest center $c'(x)$. 
Finally, each weight has weight $\poly(nd\Delta)$ and thus can be approximated to $(1+\eps)$-multiplicative factor by storing the exponent, using $\O{\log\log(nd\Delta)}$ bits of space. 
Therefore, the output of the algorithm is encoded using $\O{dk\log(n\Delta)}+|X|\cdot\polylog\left(k,\frac{1}{\eps},\log(nd\Delta),\log\frac{1}{\delta}\right)$ total bits of space.
\end{proof}

\subsection{Space-Efficient Clustering in the Streaming Model}
In this section, we show how to obtain a global encoding of the coresets inspired by the efficient encoding from the previous section.
Then, we implement our framework in \secref{sec:frame:space:front} and achieve our main algorithm for space-efficient $(k,z)$-clustering on insertion-only data streams.

The main shortcoming of immediately applying the previous efficient encoding to each node of a merge-and-reduce tree is that the constant-factor approximation requires $\O{kd\log(n\Delta)}$ bits of storage per efficient encoding, and there can be $\polylog(\log(n\Delta))$ such efficient encodings due to the height of the merge-and-reduce tree on a data stream produced by online sensitivity sampling. 
Instead, we provide a single constant-factor approximation to the global dataset and show that the error of maintaining such a global constant-factor approximation does not compound too much over the course of the stream. 

The algorithm appears in full in \algref{alg:cluster:insert:stream}, which implements our space-efficient framework in \algref{alg:framework:space} using the global encoding scheme.

\begin{algorithm}[!htb]
\caption{$(k,z)$-Clustering on Insertion-Only Streams}
\alglab{alg:cluster:insert:stream}
\begin{algorithmic}[1]
\Require{Data set $X=\{x_1,\ldots,x_n\}\subset\mathbb{R}^d$ that arrives as a data stream, $\eps\in(0,1)$, number of clusters $k$, parameter $z\ge 1$}
\Ensure{$(1+\eps)$-coreset for $(k,z)$-clustering}
\State{$Z\gets\emptyset$, $\lambda\gets\O{\frac{k}{\eps^2}\cdot\log k\log n}$}
\For{each $t\in[n]$}
\State{Let $\widehat{x_t}$ be the image of $x_t$ after applying a JL transform into $\O{\log n}$ dimensions}
\State{Use $Z\cup\{\widehat{x_t}\}$ to compute a $\O{1}$-approximation $\widehat{\sigma(x_t)}$ to the online sensitivity of $x_t$}
\State{$p(x_t)\gets\min(1,\lambda\cdot\widehat{\sigma(x_t)})$}
\State{With probability $p(x_t)$, sample $x_t$ into a stream $\calS'$ with weight $\frac{1}{p(x_t)}$}
\State{Let $Z$ be the running output of merge-and-reduce on $\calS'$ using the efficient encoding for coreset construction in \algref{alg:coreset:cluster} with a \emph{global} constant-factor approximation, for accuracy $\frac{\eps}{\poly(\log\log n\Delta)}$ and failure probability $\frac{1}{\poly\left(\frac{1}{\eps},\log(n\Delta)\right)}$} 
\Comment{Detailed implementation of \algref{alg:framework:space}}
\EndFor
\State{\Return $Z$}
\end{algorithmic}
\end{algorithm}
We first show that given a partition $X_1\sqcup\ldots\sqcup X_m$ of $X$, we can apply \algref{alg:coreset:cluster} and although the output $X'_i$ from our efficient encoding will no longer necessarily be a strong coreset for each corresponding $X_i$, the resulting error will only be an additive $\eps'\cdot\Cost(X,C)$, which suffices for our purposes. 
Specifically, we will have $\eps'=\frac{\eps}{\poly(\log\log(nd\Delta))}$ and ultimately have $m=\poly(\log\log(nd\Delta))$. 
\begin{lemma}
\lemlab{lem:each:coreset:cluster}
Let $X_1\sqcup\ldots\sqcup X_m=X$ be a partition of $X$ for $m=\poly(\log\log(nd\Delta))$. 
Let $C'$ be a set of $k$ centers that is a set $C'$ of $k$ centers that is a constant-factor approximation to $(k,z)$-clustering on $X$. 
For each $i\in[m]$, let $X'_i$ be the set of rounded points corresponding to $X_i$ from \algref{alg:coreset:cluster}. 
Then for all $i\in[m]$ and all $C\subset[\Delta]^d$ with $|C|\le k$, 
\[|\Cost(X'_i,C)-\Cost(X_i,C)|\le\frac{\eps}{\poly(\log\log nd\Delta)}\cdot\Cost(X,C).\]
\end{lemma}
\begin{proof}
Note that by \lemref{lem:x:to:xprime:cluster} with accuracy $\frac{\eps}{\poly(\log\log nd\Delta)}$, we have
\[|\Cost(X',C)-\Cost(X,C)|\le\frac{\eps}{\poly(\log\log nd\Delta)}\cdot\Cost(X,C).\]
Since 
\[|\Cost(X'_i,C)-\Cost(X_i,C)|\le\sum_{i=1}^m|\Cost(X'_i,C)-\Cost(X_i,C)|\le\eps\cdot\Cost(X,C),\]
then the claim follows. 
\end{proof}

It remains to give the full guarantees of \algref{alg:cluster:insert:stream} by showing correctness of the global encoding and in particular, that there is no compounding error across the $\poly(\log\log(nd\Delta))$ iterations. 
\begin{theorem}
\thmlab{thm:cluster:stream}
Given a set $X$ of $n$ points on $[\Delta]^d$, let $f\left(n,d,\Delta,k,\frac{1}{\eps},z\right)$ be the number of points of a coreset construction with weights $[1,\poly(nd\Delta)]$ for $(k,z)$-clustering.  
Then \algref{alg:cluster:insert:stream} outputs a $(1+\eps)$-coreset of $X$ and uses $\O{dk\log(n\Delta)}+f\left(n,d,\Delta,k,\frac{\polylog(\log(nd\Delta))}{\eps},z\right)\cdot\polylog\left(\frac{1}{\eps},\log(nd\Delta)\right)$ bits of space.
\end{theorem}
\begin{proof}
We first show that with high probability, the following invariants are satisfied at each time:
\begin{enumerate}
\item
$\widehat{\sigma(x_1)},\ldots,\widehat{\sigma(x_t)}$ are constant-factor approximations to the online sensitivities $\sigma(x_1),\ldots,\sigma(x_t)$ with respect to $(k,z)$-clustering
\item
At most $\poly\left(k,\frac{1}{\eps},\log(nd\Delta)\right)$ points have been sampled into $\calS'$
\item
$Z$ at time $t$ is a $(1+\eps)^2$-coreset to $X_t:=\{x_1,\ldots x_t\}$
\end{enumerate}
We prove correctness by induction on $t$. 
The online sensitivity of the first point is $1$ and it will be correctly computed by the data structure. 
Thus we have that after the first step, we have $\calS'=Z=\{x_1\}$ and the base case is complete. 

Let $\calE_t$ be the event that the invariants hold at time $t$.  
Conditioning on $\calE_{t-1}$ and in particular the correctness of $Z$ at time $t-1$ being a $(1+\eps)$-coreset of $X_{t-1}:=\{x_1,\ldots,x_{t-1}\}$, we have that $Z\cup\{x_t\}$ is a strong coreset of $X_t:=\{x_1,\ldots,x_t\}$. 
Thus we can use $Z\cup\{x_t\}$ to compute a constant-factor approximation to the online sensitivity $\sigma(x_t)$ of $x_t$. 
By \thmref{thm:jl}, it follows that with high probability, we can instead use $Z\cup\{\widehat{x_t}\}$ to compute a constant-factor approximation $\widehat{\sigma(x_t)}$ to the online sensitivity $\sigma(x_t)$. 
Then by \thmref{thm:online:sens:cluster}, $\calS'$ after time $t$ will be a $(1+\eps)$-coreset for $X_t$ with high probability. 

By \thmref{thm:online:sens:cluster} through \thmref{thm:total:online:sens}, we have that by sampling points with probabilities that are constant-factor approximations to their online sensitivities, then the total number of sampled points into $\calS'$ is $\O{\frac{dk^2}{\eps^2}\log^3\frac{n\Delta}{\eps}}$ with high probability. 
Thus, $\PPr{\calE_t}\ge 1-\frac{1}{\poly(n)}$. 

Observe that merge-and-reduce will be correct if $\calS'$ is a stream with length at most $k\cdot\poly\left(\frac{1}{\eps},\log(nd\Delta)\right)$, since we set the failure probability to be $\frac{1}{\poly\left(\frac{1}{\eps},\log(n\Delta)\right)}$ and merge-and-reduce will consider at least $k$ stream updates before applying a new coreset construction. 
Thus conditioned on (1) the correctness of $\calS'$ after time $t$ being a $(1+\eps)$-coreset for $X_t$ and (2) the correctness of merge-and-reduce on $\calS'$, we have that $Z$ after time $t$ will be a $(1+\eps)$-coreset for $\calS'$, and thus a $(1+\eps)^2$-strong coreset for $X_t$, which completes the induction. 
That is, $\PPr{\calE_t\,\mid\,\calE_1,\ldots,\calE_{t-1}}\ge 1-\frac{1}{\poly(n)}$. 
The correctness guarantee then follows by rescaling $\eps$ and a union bound over all $t\in[n]$. 

To analyze the space complexity, note that by the guarantees of the invariants, the total number of sampled points into $\calS'$ is $\O{\frac{dk^2}{\eps^2}\log\frac{nd\Delta}{\eps}}$. 
However, $\calS'$ is not maintained explicitly, but instead given as input to the merge-and-reduce subroutine, which uses the efficient encoding for coreset construction given in \algref{alg:coreset:cluster} with accuracy $\frac{\eps}{\poly(\log\log(nd\Delta))}$ and failure probability $\frac{1}{\poly\left(\frac{1}{\eps},\log(nd\Delta)\right)}$. 
Thus by \lemref{lem:efficient:coreset:cluster}, the total representation of $Z$ uses $\O{dk\log(n\Delta)}+\frac{dk^2}{\eps^2}\cdot2^{2z}\cdot\polylog\left(\frac{1}{\eps},\log(n\Delta)\right)$ bits of space.
\end{proof}

\section{Subspace Embeddings}
\applab{sec:subspace:embed}
In this section, we provide our one-pass streaming algorithm on insertion-only streams for $L_p$ subspace embeddings using $\mathcal{O}_{k,d,\eps}(1)$ words of space. 
Recall that in te problem of constructing $L_p$ subspace embeddings, the goal is to produce a matrix $\bM\in\mathbb{R}^{m\times d}$ such that for an accuracy parameter $\eps \in (0,1)$ and all vectors $\bx\in\mathbb{R}^d$, we have 
\[(1-\eps)\|\bA\bx\|_p\le\|\bM\bx\|_p\le(1+\eps)\|\bA\bx\|_p.\]
$L_p$ subspace embeddings serve as a foundational tool for data compression by maintaining the essential geometric structure of high-dimensional datasets. 
This preservation is vital for various downstream tasks, including dimensionality reduction, compressed sensing, and linear regressions.
Hence, there is a long line of work studying $L_p$ subspace embeddings~\cite{DrineasMM06a,DrineasMM06b,Sarlos06,DasguptaDHKM09,ClarksonW09,CohenP15,CohenMP20,BravermanDMMUWZ20,ParulekarPP21,MahabadiWZ22,MeyerMMWZ22,MuscoMWY22,CherapanamjeriSWZ23,MeyerMMWZ23,WoodruffY23}. 

We first define the leverage scores, which describes the importance of a row for $L_2$ subspace embeddings. 
 
\begin{definition}[Leverage scores]
\deflab{def:leverage:score}
The \emph{leverage score} of a row $\ba_i\in\mathbb{R}^d$ of a matrix $\bA\in\mathbb{R}^{n\times d}$ is  
\[\max_{\bx:\|\bA\bx\|_2=1}\langle\ba,\bx\rangle^2 = \ba_i^\top(\bA^\top\bA)^{-1}\ba_i.\]
\end{definition}
For the purpose of constructing $L_p$ subspace embedding, \cite{CohenP15} generalized the above definition to $L_p$ Lewis weights, defined as follows:

\begin{definition}[$L_p$ Lewis weights]
\deflab{def:lewis:weight}
The \emph{Lewis weight} of a row $\ba_i\in\mathbb{R}^d$ of a matrix $\bA\in\mathbb{R}^{n\times d}$ is the $i$-th diagonal entry of a diagonal weight matrix $\bW$ such that for all $i\in[n]$
\[W_{i,i}=\ell_i(\bW^{\frac{1}{2}-\frac{1}{p}}\bA),\]
where $\ell_i(\bW^{\frac{1}{2}-\frac{1}{p}}\bA)$ denotes the leverage score of the $i$-th row of the matrix $\bW^{\frac{1}{2}-\frac{1}{p}}\bA$. 
\end{definition}

$L_p$ Lewis weights may be viewed as leverage scores computed after a density reweighting that balances the rows so that preserving $\|\bA\bx\|_p$ reduces to preserving the Euclidean norm $\|\bA'\bx\|_2$ for a suitably rescaled matrix $\bA'$~\cite{CohenP15}.  
Compared with alternative row-importance measures such as $L_p$ leverage scores~\cite{DasguptaDHKM09} or $L_p$ sensitivities~\cite{BravermanDMMUWZ20,ChenD21}, Lewis weights achieve optimal row-sampling complexity for $L_p$ embeddings.  
For the special case $p=2$, it is known that oblivious subspace embeddings with $\O{\frac{d}{\eps^2}}$ rows exist~\cite{Sarlos06}.  
By combining these methods,we have the following guarantee.

\begin{theorem}
\thmlab{thm:lewis:weights}
\cite{Sarlos06,CohenP15,WoodruffY23}
Let $\bA\in\mathbb{R}^{n\times d}$, $\eps\in(0,1)$, and $p\ge 1$.  
Define
\[
f\left(d,\frac{1}{\eps},p\right)=\begin{cases}
\frac{d}{\eps^2}\log\frac{d}{\eps}\polylog\!\left(\frac{d}{\eps}\right), & p\in[1,2),\\[6pt]
\O{\frac{d}{\eps^2}}, & p=2,\\[6pt]
\frac{d^{p/2}}{\eps^2}\cdot\polylog\!\left(\frac{d}{\eps}\right), & p>2.     
\end{cases}
\]
There exists a coreset construction for $L_p$ subspace embedding consisting of 
$f\!\left(d,\frac{1}{\eps},p\right)$ rows, where each entry is bounded by $\|\bA\|_\infty\cdot\poly(n)$.
\end{theorem}

We now extend the notion of Lewis weights to the online setting.

\begin{definition}[Online $L_p$ Lewis weights]
\deflab{def:online:lewis:weight}
For each $t \in [n]$, the online $L_p$ \emph{Lewis weight} of a row $\ba_t\in\mathbb{R}^d$ of a matrix $\bA\in\mathbb{R}^{n\times d}$ is the Lewis weight of $\ba_i$ with respect to the prefix matrix $\bA_t=\ba_1\circ\ldots\circ\ba_t$, i.e., the submatrix of $\bA$ consisting of the first $t$ rows of $\bA$. 
\end{definition}

We recall bounds on the total mass of online Lewis weights.

\begin{theorem}
\thmlab{thm:total:online:lewis}
\cite{BravermanDMMUWZ20,WoodruffY23}
Let $\bA=\{\ba_1,\ldots,\ba_n\}\in\mathbb{R}^{n\times d}$ and let $\sigma(\ba_t)$ denote the online $L_p$ Lewis weight of row $\ba_t$ for $t\in[n]$, with $p\ge 1$.  
Then, for $p\in[1,2]$,
\[
\sum_{t=1}^n \sigma(\ba_t)
= d\cdot \polylog(n\kappa),
\]
and for $p>2$,
\[
\sum_{t=1}^n \sigma(\ba_t)
= d^{p/2}\cdot \polylog(n\kappa),
\]
where $\kappa$ denotes the online condition number.
\end{theorem}

Finally, we state the guarantees provided by sampling according to online Lewis weights.

\begin{theorem}[Online Lewis weight sampling]
\thmlab{thm:online:lewis:embedding}
\cite{WoodruffY23}
Given a stream of rows $\ba_1,\ldots,\ba_n\in\mathbb{R}^{n\times d}$, we sample row $\ba_t$ with probability $p_t \ge \min(1,\gamma\cdot\sigma(\ba_t))$, where $\sigma(\ba_t)$ is its online Lewis weight and  $\gamma=\O{\frac{\log n}{\eps^2}}$, and rescale $\ba_t$ by a factor $\frac{1}{p_t}$ if sampled.  
Then, with high probability, the resulting weighted sample forms a $(1+\eps)$-coreset for $L_p$ subspace embedding whose size is at most
\[
g\left(n,d,\frac{1}{\eps},\kappa,p\right)=\begin{cases}
\frac{d}{\eps^2}\cdot\polylog(n\kappa), & p\in[1,2],\\[6pt]
\frac{d^{p/2}}{\eps^2}\cdot\poly(\log^{p/2}(n\kappa)), & p>2.     
\end{cases}
\]
\end{theorem}

We use standard approaches, e.g., Lemma 4.1 in \cite{ClarksonW09} to upper bound the magnitude of the logarithm of any nonzero singular value of a matrix with bounded integer entries. 
This in turn upper bounds the logarithm of the online condition number. 
\begin{lemma}
Suppose $\bA\in\mathbb{R}^{n\times d}$ has integer entries bounded in magnitude by $M$. 
Then for any nonzero singular value $\sigma_i$ of $\bA$, we have $|\log\sigma_i|=\O{d\log(nM)}$.
\end{lemma}
\begin{proof}
We have that the characteristic polynomial of $\bA^\top\bA$ is $p(x)=\det(x\cdot\mathbb{I}_d-\bA^\top\bA)$. 
Note that if $\bA$ has rank $r$, then $p(x)=x^{d-r}\prod_{i=1}^r(x-\lambda_i)$, where $\lambda_i$ is the $i$-th eigenvalue of $\bA^\top\bA$, writing $\lambda_1\ge\lambda_2\ge\ldots\ge\lambda_r$. 
Since all entries of $\bA^\top\bA$ are integers, then the coefficients of $p(x)$ are also integers. 
Moreover, since the eigenvalues of $\bA^\top\bA$ are non-negative, then $\prod_{i=1}^r\lambda_i\ge 1$. 
Because all entries of $\bA^\top\bA$ are at most $\poly(ndM)$ in magnitude, then 
\[\lambda_i\le\lambda_1\le\|\bA^\top\bA\|_F\le\poly(ndM).\]
Since $\prod_{i=1}^r\lambda_i\ge 1$, then we have that $\lambda_i\ge\frac{1}{\poly(ndM)^d}$. 
Since $\sigma_i^2=\lambda_i$, then $|\log\sigma_i|=\O{d\log(nM)}$.
\end{proof}

We next recall the following definition of a well-conditioned basis.
\begin{definition}[Well-conditioned basis]
Given a matrix $\bA\in\mathbb{R}^{n\times d}$ of rank $r$, let $p\in[1,\infty)$ and $q$ be its dual norm, so that $\frac{1}{p}+\frac{1}{q}=1$. 
Then matrix $\bU\in\mathbb{R}^{n\times d}$ is an $(\alpha,\beta,p)$-well-conditioned basis for the column space of $\bA$ if:
\begin{enumerate}
\item 
The column space of $\bA$ is the column space of $\bU$
\item
$\sum_{i\in[n],j\in[d]} U_{i,j}^p\le\alpha^p$
\item
For all $\bx\in\mathbb{R}^d$, we have $\|\bz\|_q\le\beta\|\bU\bz\|_p$
\end{enumerate}
\end{definition}
One such construction of a well-conditioned basis gives the following properties:
\begin{theorem}
\cite{DasguptaDHKM09}
\thmlab{thm:well:cond:basis}
Given a matrix $\bA\in\mathbb{R}^{n\times d}$ of rank $r$, let $p\in[1,\infty)$ and $q$ be its dual norm, so that $\frac{1}{p}+\frac{1}{q}=1$. 
There exists an $(\alpha,\beta,p)$-well-conditioned basis $\bU$ for the column space of $\bA$ such that:
\begin{itemize}
\item 
If $p<2$, then $\alpha=r^{\frac{1}{p}+\frac{1}{2}}$ and $\beta=1$
\item 
If $p=2$, then $\alpha=\sqrt{r}$ and $\beta=1$
\item 
If $p>2$, then $\alpha=r^{\frac{1}{p}+\frac{1}{2}}$ and $\beta=r^{\frac{1}{q}-\frac{1}{2}}$
\end{itemize}
Moreover, $\bU$ can be computed in time $\O{ndr+nr^2\log n}$. 
\end{theorem}
For completeness, we briefly describe the construction of the well-conditioned basis of \thmref{thm:well:cond:basis}, given by \cite{DasguptaDHKM09}. 
Given a QR decomposition $\bA=\bQ\bR$, so that $\bQ$ is any $n\times r$ matrix that is an orthonormal basis for the span of $\bA$ and $\bR$ is an $r\times d$ matrix, define the set $S=\{\bx\in\mathbb{R}^d\,\mid\,\|\bQ\bx\|_p\le1\}$ and the $r\times r$ matrix $\bF$ so that $\calE_{LJ}=\{\bx\in\mathbb{R}^r\,\mid\,\bx^\top\bF\bx\le 1\}$ is the L\"{o}wner-John ellipsoid of $S$, let $\bG\in\mathbb{R}^{d\times r}$ be the full rank and upper triangular matrix such that $\bF=\bG^\top\bG$. 
Then $\bU:=\bQ\bG^{-1}$ is the desired $(\alpha,\beta,p)$-well-conditioned basis for the column span of $\bA$. 

We now define the $L_p$ sensitivity of a row of a matrix, for the purposes of $L_p$ subspace embeddings. 
\begin{definition}[$L_p$ sensitivity]
For a matrix $\bA\in\mathbb{R}^{n\times d}$, the \emph{$L_p$ sensitivity} of row $\ba_t$ is the quantity
\[\max_{\by\in\mathbb{R}^d: \|\bA\by\|_p=1}\left\lvert\langle\ba_t,\by\rangle\right\rvert^p.\]
\end{definition}
Finally, we recall the following upper bound on the total $L_p$ sensitivity for $L_p$ subspace embedding. 
\begin{lemma}[Upper bounds on the sum of the $L_p$ sensitivities]
\cite{CohenP15,ChenD21}
\lemlab{lem:total:lp:sens}
For a matrix $\bA\in\mathbb{R}^{n\times d}$, let $s(\ba_t)$ denote the $L_p$ sensitivity of the $t$-th row of $\bA$ for $L_p$ subspace embedding. 
Then $\sum_{t=1}^n s(\ba_t)\le d$ for $p\le 2$ and $\sum_{t=1}^n s(\ba_t)\le d^{p/2}$ for $p>2$. 
\end{lemma}

\subsection{Efficient Encoding for Coreset Construction for \texorpdfstring{$L_p$}{Lp} Subspace Embedding}
\applab{sec:lp:encoding}
Given a matrix $\bA$, which can be viewed as either the original matrix or a set of reweighted rows that forms a coreset of some underlying matrix, we first acquire a constant-factor approximation $\bM$ for $L_p$ subspace embedding on $\bA$. 
We then use the existence of a well-conditioned basis to compute a deterministic preconditioner $\bP\in\mathbb{R}^{d\times d}$, so that $\bM\bP^{-1}$ has condition number $\poly(d)$. 
Now for each row $\ba_t$ of $\bA$, let $\bb'_t$ be $\ba_t\bP$ with each coordinate rounded to a power of $(1+\eps')$ for $\eps'=\frac{\poly(\eps)}{\poly(d^p,\log(nd))}$. 
We then store the exponent of each rounded coordinate of $\bB'$, as well as $\bM$. 
We state the algorithm in \algref{alg:coreset:embedding}. 

\begin{algorithm}[!htb]
\caption{Efficient Encoding for Coreset Construction for $L_p$ Subspace Embedding}
\alglab{alg:coreset:embedding}
\begin{algorithmic}[1]
\Require{Matrix $\bA\subset\{-M,\ldots,-1,0,1,\ldots,M\}^{n\times d}$, accuracy parameter $\eps\in(0,1)$, failure probability $\delta\in(0,1)$}
\Ensure{$(1+\eps)$-coreset for $L_p$ subspace embedding}
\State{$\eps'\gets\frac{\poly(\eps)}{\poly(d^p,\log(nd\Delta))}$}
\State{Find a matrix $\bM$ that is a constant-factor $L_p$ subspace embedding on $\bA$}
\Comment{\thmref{thm:lewis:weights}}
\State{Compute a deterministic preconditioner $\bP$ of $\bM$ using well-conditioned bases, such that $\bM\bP^{-1}$ has condition number $\poly(d)$}
\For{each row $\ba\in\bA$}
\State{Let $\bb'$ be $\ba\bP$ with each coordinate rounded to a power of $(1+\eps')$}
\State{$\bB'\gets\bB'\circ\bb'$, storing the \emph{exponent} for each entry of $\bb'$}
\EndFor
\State{\Return $(\bM,\bB')$, from which estimate $\bA'$ can be constructed}
\end{algorithmic}
\end{algorithm}
Given a constant-factor subspace embedding $\bM$, we now show that the matrix $\bB'$ acquired by rounding each row of $\bA\bP$ can be used to construct a matrix $\bA'$ that is a strong coreset of $\bA$. 
\begin{lemma}
\lemlab{lem:x:to:xprime:embedding}
Let $\eps\in\left(0,\frac{1}{2}\right)$ and let $\bA'=\bB'\bP^{-1}$, where $\bB'$ is the output of \algref{alg:coreset:embedding} and $\bP$ is the deterministic preconditioner of $\bM$. 
Then for all $\bx\in\mathbb{R}^d$,
\[(1-\eps)\|\bA\bx\|_p^p\le\|\bA'\bx\|_p\le(1+\eps)\|\bA\bx\|_p^p.\]
\end{lemma}
\begin{proof}
Without loss of generality, we assume that $\bA$ is full rank. 
Note that we can write $\bA\bx=\bA\bP\bP^{-1}\bx$ and thus by rewriting $\by=\bP^{-1}\bx$, it suffices to show that for all $\by\in\mathbb{R}^d$
\[(1-\eps)\|\bA\bP\by\|_p^p\le\|\bA'\bP\by\|_p^p\le(1+\eps)\|\bA\bP\by\|_p^p.\]
We remark that if $\bA$ is not full rank, the task would be to show claim for all $\by$ in the column span of $\bP^{-1}$. 

Let $\bB=\bA\bP$ and observe that \algref{alg:coreset:embedding} rounds all coordinates of each row $\bb_t=\ba_t\bP$ to their closest power of $(1+\eps')$. 
Then we have 
\[\left\lvert\langle\bb_t,\by\rangle^p-\langle\bb'_t,\by\rangle^p\right\rvert\le\eps'\cdot\max_{\by\in\mathbb{R}^d,\|\by\|_p=1}\left\lvert\langle\bb_t,\by\rangle\right\rvert^p.\]
For each $t\in[n]$, let $s(\bb_t)$ denote the sensitivity of $\bb_t$ with respect to $\bB$, so that
\[s(\bb_t)=\max_{\by\in\mathbb{R}^d}\frac{\left\lvert\langle\bb_t,\by\rangle\right\rvert^p}{\|\bB\by\|_p^p}.\]
Recall that $\bB=\bA\bP$ and $\bP$ is a preconditioner for $\bM$, so that $\bM\bP$ is well-conditioned. 
That is, for any unit vector $\by\in\mathbb{R}^d$, we have
\[\frac{1}{d^{pC}}\cdot\|\bM\bP\by\|_p^p\le\|\by\|_p^p\le d^{pC}\cdot\|\bM\bP\by|_p^p,\]
for a fixed constant $C>0$. 
Since $\bM$ is a constant-factor subspace embedding for $\bA$, then we certainly have 
\[\frac{1}{d^{p(C+1)}}\cdot\|\bA\bP\by\|_p^p\le\|\by\|_p^p\le d^{p(C+1)}\cdot\|\bA\bP\by|_p^p.\]
Thus it follows that 
\[\frac{1}{d^{p(C+1)}}\cdot\|\bB\by\|_p^p\le\|\by\|_p^p\le d^{p(C+1)}\cdot\|\bB\by|_p^p.\]
Therefore, 
\[\max_{\by\in\mathbb{R}^d}\left\lvert\langle\bb_t,\by\rangle\right\rvert^p\le s(\bb_t)\cdot d^{p(C+1)},\]
so that
\[\left\lvert\|\langle\bb_t,\by\rangle^p-\langle\bb'_t,\by\rangle^p\right\rvert\le\eps'\cdot s(\bb_t)\cdot d^{p(C+1)}.\]
Then by triangle inequality, we have
\[\left\lvert\|\bB\by\|_p^p-\|\bB'\by\|_p^p\right\rvert\le\sum_{t=1}^n\left\lvert\|\langle\bb_t,\by\rangle-\langle\bb'_t,\by\rangle\right\rvert^p\le\eps'\cdot\sum_{t=1}^n s(\bb_t)\cdot d^{p(C+1)}.\]
Since the sum of $L_p$ sensitivities is at most $d$ by \lemref{lem:total:lp:sens}, then 
\[\left\lvert\|\bB\by\|_p^p-\|\bB'\by\|_p^p\right\rvert\le\eps'\cdot
d^{p(C+2)}.\]
The desired claim then follows by the setting of $\eps'=\frac{\poly(\eps)}{\poly(d^p,\log(nd\Delta))}$, recalling that $\bB\by=\bA\bP\bP^{-1}\bx$ and $\bB'\by=\bA'\bP\bP^{-1}\bx$. 
\end{proof}
We now give the full guarantees of \algref{alg:coreset:embedding}, which gives an efficient encoding of an input matrix $\bA$. 
\begin{lemma}
\lemlab{lem:efficient:coreset:embedding}
Let $\bA\in\mathbb{R}^{m\times d}$ be a coreset construction with entries in $\{0\}\cup\left[\frac{1}{\poly(n)},\poly(n)\right]$. 
Let 
\[f\left(d,\frac{1}{\eps},p\right)=\begin{cases}
\frac{d}{\eps^2}\log\frac{d}{\eps}\polylog\left(\frac{d}{\eps}\right),\qquad& p\in[1,2)\\
\O{\frac{d}{\eps^2}},\qquad& p=2\\
\frac{d^{p/2}}{\eps^2}\cdot\polylog\left(\frac{d}{\eps}\right),\qquad&p>2     
\end{cases}.\]
Then $\bA'$ is a $(1+\eps)$-coreset for $\bA$ for $L_p$ subspace embedding that uses 
$\O{f\left(d,1,p\right)}\cdot (d\log n)+md\cdot\polylog\left(d,\frac{1}{\eps},\log(nd\kappa),\log\frac{1}{\delta}\right)$ bits of space. 
\end{lemma}
\begin{proof}
Firstly, note that \algref{alg:coreset:embedding} outputs a pair $(\bM,\bB')$ that encodes $\bB'$ using $\bM$. 
By \lemref{lem:x:to:xprime:embedding}, we have that for all $\bx\in\mathbb{R}^d$, 
\[(1-\eps)\|\bA\bx\|_1\le\|\bA'\bx\|_1\le(1+\eps)\|\bA\bx\|_1,\]
where $\bA'=\bB'\bP^{-1}$. 
Thus, $\bA'$ is a coreset for $\bA$. 
 
It remains to analyze the space complexity of $(\bM,\bB')$. 
Since $\bM$ is a constant-factor $L_p$ subspace embedding, then by \thmref{thm:lewis:weights}, $\bM$ can be represented using $\O{f(d,1,p)}\cdot d\log(n)$ bits of space. 
Furthermore, each offset $\bb'$ in $\bB'$ is encoded using $\O{d\log\left(\frac{1}{\eps}+\log(nd\kappa)\right)}$ bits of space due to storing the $d$ coordinates of the offset rounded to the power of $(1+\eps')$. 
Specifically, we store the exponent of each offset after the rounding, rather than the explicit coordinates. 
Therefore, the output of the algorithm is encoded using $\O{f\left(d,1,p\right)}\cdot (d\log n)+md\cdot\polylog\left(d,\frac{1}{\eps},\log(nd\kappa),\log\frac{1}{\delta}\right)$. 
\end{proof}

\subsection{Subspace Embedding in the Streaming Model}
\applab{sec:lp:space}
In this section, we show how to use a global encoding combined with the efficient encoding from the previous section to achieve our main algorithm for $L_p$ subspace embeddings in the row-arrival model, using our space-efficient framework in \secref{sec:frame:space:front}. 

Similar to $(k,z)$-clustering, the main downfall of immediately applying the previous efficient encoding to each node of a merge-and-reduce tree is that the constant-factor approximation requires $\O{d^2\log(nd)}$ bits of storage per efficient encoding, and there can be $\polylog(\log(nd))$ such efficient encodings due to the height of the merge-and-reduce tree on a data stream produced by online Lewis weight sampling. 
As before, we circumvent this issue by showing that we can instead consider a single constant-factor approximation to the global dataset. 
We then show that the error of maintaining such a global constant-factor approximation does not compound too much over the course of the stream. 

We remark that in our algorithm, we apply the online coreset construction for $L_p$ subspace embedding as a black box \cite{WoodruffY23}.
As we mentioned earlier, in online coreset sampling, if we use the previous sketch maintained by the algorithm to compute the future sampling probabilities, the estimation will depend on whether the previous rows are sampled, and so standard concentration inequalities do not apply.
To avoid this dependency, \cite{WoodruffY23} kept a separate constant-factor $L_2$ subspace embedding sketch and used it to compute the future online sampling probabilities of the $L_p$ subspace embedding sketch, decoupling the randomness of estimating the online sampling probabilities and that of sampling the rows.
We observe that by passing an online $L_2$ subspace embedding sketch (e.g., by online row sampling \cite{CohenMP20}) to the merge-and-reduce structure, we obtain a constant-factor approximation with $d \poly(\log \log (n\kappa))$ rows.
Thus, we can store this sketch by our global efficient encoding in $\O{d}$ words, which does not exceed our space budget.
We give the algorithm in full in \algref{alg:embedding:insert:stream}, which applies our space-efficient framework in \algref{alg:framework:space}. 

\begin{algorithm}[!htb]
\caption{$L_p$-subspace embedding on insertion-only streams}
\alglab{alg:embedding:insert:stream}
\begin{algorithmic}[1]
\Require{Matrix $\bA=\{\ba_1,\ldots,\ba_n\}\in\mathbb{R}^{n\times d}$ that arrives as a data stream, $\eps\in(0,1)$, parameter $p\ge 1$}
\Ensure{$(1+\eps)$-coreset for $L_p$-subspace embedding}
\State{$\bZ, \bZ'\gets\emptyset$, $\lambda\gets\O{\frac{d}{\eps^2}\cdot\log d\log n}$}
\For{each $t\in[n]$}
\State{Use $\bZ'\cup\{\widehat{\ba_t}\}$ to compute a $\O{1}$-approximation $\widehat{\sigma(\ba_t)}$ to the online Lewis weight of $\ba_t$}
\State{$p(\ba_t)\gets\min(1,\lambda\cdot\widehat{\sigma(\ba_t)})$}
\State{With probability $p(\ba_t)$, sample $\frac{1}{p(\ba_t)}\cdot\ba_t$ into a stream $\calS'$}
\State{Let $\bZ$ be the running output of merge-and-reduce on $\calS'$ using the efficient encoding for coreset construction in \algref{alg:coreset:embedding} with a \emph{global} constant-factor embedding, for accuracy $\frac{\eps}{\poly(\log\log n \kappa)}$ and failure probability $\frac{1}{\poly\left(\frac{1}{\eps},\log(n \kappa)\right)}$} \Comment{Detailed implementation of \algref{alg:framework:space}}
\State{Maintain a separate constant-factor $L_2$ subspace embedding sketch $\bZ'$ with efficient encoding}
\EndFor
\State{\Return $\bZ$}
\end{algorithmic}
\end{algorithm}
We first show that given disjoint matrices $\bQ_1,\ldots,\bQ_m$ of $\bA$, we can apply \algref{alg:coreset:embedding} and although the output $\bQ'_i$ from our efficient encoding will no longer necessarily be a strong coreset for each corresponding $\bQ_i$, the resulting error will only be an additive $\eps'\cdot\|\bA\bx\|_p^p$ for a fixed $\bx\in\mathbb{R}^d$, which suffices for our purposes. 
Specifically, we will set $\eps'=\frac{\eps}{\poly(\log\log(n\kappa))}$ and $m=\poly(\log\log(n\kappa))$, provided the entries of $\bA$ are integers bounded in magnitude by $\poly(nd)$. 
\begin{lemma}
\lemlab{lem:each:coreset:embedding}
Let $\bQ_1\circ\ldots\circ\bQ_m=\bA$ be a partition of $\bA$. 
Let $\bM$ be a constant-factor $L_p$ subspace embedding on $\bA$. 
For each $i\in[m]$, let $\bQ'_i$ be the resulting matrix after applying the inverse of the preconditioner to the rounded matrix resulting from $\bQ_i$. 
Then for all $i\in[m]$ and all $\bx\in\mathbb{R}^d$, 
\[\left\lvert\|\bQ'_i\bx\|_p^p-\|\bQ_i\bx\|_p^p\right\rvert\le\frac{\eps}{\poly(\log\log n\kappa)}\cdot\|\bA\bx\|_p^p.\]
\end{lemma}
\begin{proof}
Note that by \lemref{lem:x:to:xprime:embedding} with accuracy $\frac{\eps}{\poly(\log\log n\kappa)}$, we have
\[\left\lvert\|\bA'\bx\|_p^p-\|\bA\bx\|_p^p\right\rvert\le\frac{\eps}{\poly(\log\log n\kappa)}\cdot\|\bA\bx\|_p^p.\]
Since 
\[\left\lvert\|\bQ'_i\bx\|_p^p-\|\bQ_i\bx\|_p^p\right\rvert\le\sum_{i=1}^m\left\lvert\|\bQ'_i\bx\|_p^p-\|\bQ_i\bx\|_p^p\right\rvert=\left\lvert\|\bA'\bx\|_p^p-\|\bA\bx\|_p^p\right\rvert,\]
then the claim follows. 
\end{proof}
We now give the full guarantees of \algref{alg:embedding:insert:stream} by showing correctness of the global encoding and specifically, that there is no compounding error across the $\poly(\log\log(nd))$ iterations. 
We remark that our update time in each step is dominated by reconstructing the sketch and computing the online sampling probability, and so the update time is $\poly(d)$, where the power of $d$ depends on the matrix-multiplication exponential.
\begin{theorem}
\thmlab{thm:subspace:per:update}
Given a matrix $\bA$ of $n$ rows on $[-M,\ldots,-1,0,1,\ldots,M]^d$, with $M=\poly(n)$ and online condition number $\kappa$ defined as above, there exists a one-pass streaming algorithm in the row arrival model that outputs a $(1+\eps)$-coreset of $\bA$ and uses $\O{d}\cdot f\left(d,\frac{1}{\eps},p\right)$ words of space, where
\[f\left(d,\frac{1}{\eps},p\right)=\begin{cases}
\frac{d}{\eps^2}\log\frac{d}{\eps}\polylog\left(\frac{d}{\eps}\right),\qquad& p\in[1,2)\\
\O{\frac{d}{\eps^2}},\qquad& p=2\\
\frac{d^{p/2}}{\eps^2}\cdot\polylog\left(\frac{d}{\eps}\right),\qquad&p>2     
\end{cases}.\]
The (amortized) update time of this algorithm is $\poly(d)$.
\end{theorem}
\begin{proof}
Consider \algref{alg:embedding:insert:stream}. 
We prove correctness by induction on $t$, claiming that after $j\cdot f\left(d,\frac{1}{\eps},p\right)$ samples into $\calS'$, then $\bZ$ is a $\left(1+\O{\eps}\right)^{j+1}$ coreset of $\bA_t$. 
The online Lewis weight of the first row is always $1$ and thus $\bZ=\ba_1$ after the first step, and the base case is complete. 
For the inductive hypothesis, we condition on the correctness of $\bZ$ being a $\left(1+\O{\eps}\right)^j$-coreset of $\bA_{t-1}=\ba_1\circ\ldots\circ\ba_{t-1}$ after $j-1$ merges have occured. 
Then $\bZ\circ\ba_t$ is a $\left(1+\O{\eps}\right)^j$-coreset of $\bA_t=\ba_1\circ\ldots\circ\ba_t$, so we can use $\bZ\circ\ba_t$ to compute a constant-factor approximation to the online Lewis weight $\sigma(\ba_t)$ of $\ba_t$ for a sufficiently small $\eps$ to be specified later. 
Note that merge-and-reduce will be correct if $\calS'$ is a stream with length at most $\O{g\left(n,d,\frac{1}{\eps},\kappa, p\right)}$, since the failure probability is set to be $\frac{1}{\poly\left(\frac{1}{\eps},\log(n\kappa)\right)}$ and merge-and-reduce will consider at least $f\left(d,\frac{1}{\eps},p\right)$ stream updates before applying a new coreset construction. 
Thus by \lemref{lem:each:coreset:embedding} the correctness of merge-and-reduce on $\calS'$, we have that $\bZ$ after time $t$ will be a $\left(1+\O{\eps}\right)^{j+1}$-strong coreset for $\bA_t$ after $j$ merges, which completes the induction. 
Finally, we have that by \thmref{thm:online:lewis:embedding} through \thmref{thm:total:online:lewis}, at most $\O{g\left(n,d,\frac{1}{\eps},\kappa, p\right)}$ samples will be inserted into $\calS$, so that $j\le \O{\log \frac{g\left(n,d,\frac{1}{\eps},\kappa, p\right)}{f\left(d,\frac{1}{\eps},p\right)}}\le\poly(\log \log(n\kappa))$. 
The correctness guarantee then follows by rescaling $\eps$ with $\poly(\log \log(n\kappa))$. 

To analyze the space complexity, recall that the total number of sampled rows into $\calS'$ is $n'=\O{g\left(n,d,\frac{1}{\eps},\kappa, p\right)}$, with high probability. 
We do not maintain $\calS'$, but instead feed $\calS'$ as input to the merge-and-reduce subroutine. 
Hence, the input to merge-and-reduce has size $n'$ and so 
note that each coreset in the merge-and-reduce tree has size $f\left(d,\frac{\poly(\log \log(n\kappa))}{\eps},p\right)$. 
We use the efficient encoding for coreset construction given in \algref{alg:coreset:embedding} with accuracy $\frac{\eps}{\poly(\log\log(n\kappa))}$ and failure probability $\frac{1}{\poly\left(\frac{1}{\eps},\log(n\kappa)\right)}$. 
Similar to the analysis of \lemref{lem:efficient:coreset:embedding}, we recall that total representation of $\bZ$ requires a constant-factor $L_p$ subspace embedding $\bM$ that uses $\O{f\left(d,1,p\right)}\cdot (d\log n)$ bits of space. 
We then use \lemref{lem:each:coreset:embedding} to encode each of the coresets of size $f\left(d,\frac{\poly(\log \log(n\kappa))}{\eps},p\right)$, using $d\cdot\polylog\left(d,\frac{1}{\eps},\log(nd\kappa),\log\frac{1}{\delta}\right)$ bits per row. 
Since the height of the merge-and-reduce tree is at most $\poly(\log \log(n\kappa))$, it follows that there are at most $\poly(\log \log(n\kappa))$ such coresets. 
Therefore, the total space required is 
\[\O{d\log n}\cdot f(d,1,p)+f\left(d,\frac{1}{\eps},p\right)\cdot\polylog(n')\cdot d\cdot\polylog\left(\frac{1}{\eps},\log(nd\kappa)\right).\]
in bits, which is equivalent to $\O{d}\cdot f\left(d,\frac{1}{\eps},p\right)$ words of space, as desired.
In addition, we maintain a constant-factor $L_2$ subspace embedding $\bZ'$, which is used to estimate the online Lewis weights. 
The construction of $\bZ'$ follows the same procedure, which passes the online sub-stream to a merge-and-reduce structure.
Note that each coreset in the merge-and-reduce tree now has size $f\left(d,\poly(\log \log(n\kappa)),2\right)$. 
We again use the efficient encoding for coreset construction given in \algref{alg:coreset:embedding} with accuracy $\frac{1}{\poly(\log\log(n\kappa))}$ and failure probability $\frac{1}{\poly\left(\frac{1}{\eps},\log(n\kappa)\right)}$. 
Thus, the total space required is 
\[\O{d\log n}\cdot f(d,1,2)+f\left(d,1,2\right)\cdot\polylog(n')\cdot d\cdot\polylog\left(\log(nd\kappa)\right).\]
in bits, which is dominated by the space to store the $L_p$ subspace embedding sketch.

To analyze the update time, at each step, we need to reconstruct the constant-factor $L_2$ subspace embedding from our global encoding, using $\poly(d, \polylog(\log n\kappa)$ update time by \thmref{thm:well:cond:basis}.
Then, we apply the reconstructed sketch to estimate the online Lewis weight of $\ba_t$, using $\poly(d)$ update time \cite{CohenMP20,WoodruffY23}.
We note that, whenever we merge some coreset, we need $\poly\left(d, \frac{1}{\eps}, \log\log (n\kappa)\right)$ update time to compute the global efficient encoding.
However, since we only conduct $\polylog(n\kappa)$ times such operation in the entire stream, they are amortized to lower order terms and hence are dominated by the time to compute the online Lewis weights.
\end{proof}

\section{Projection-Cost Preserving Sketches}
\applab{sec:pcp:all}
In this section, we describe our one-pass streaming algorithm on insertion-only streams for constructing projection-cost preserving (PCP) sketches that uses $\mathcal{O}_{k, \eps}\left(\poly(\log\log(n\kappa))\right) + \mathcal{O}_{d,\eps}(1)$ words of space. 
Recall that in the problem of constructing (PCP) sketches in the streaming setting, the rows of a matrix $\bA\in\mathbb{R}^{n\times d}$ are revealed sequentially, and the goal is to maintain a much smaller matrix $\bB\in\mathbb{R}^{m\times d}$ such that for every rank-$k$ orthogonal projection matrix $\bP\in\mathbb{R}^{d\times d}$,
\[
(1-\eps)\|\bA-\bA\bP\|_F^2 \le 
\|\bB-\bB\bP\|_F^2 
\le 
(1+\eps)\|\bA-\bA\bP\|_F^2,
\]
where $\|\cdot\|_F$ denotes the Frobenius norm.
A PCP sketch approximately preserves the total squared distance of the data to any $k$-dimensional subspace, thereby maintaining the geometric structure relevant to dimensionality reduction.  
As a result, PCP sketches play a central role in randomized numerical linear algebra, including applications to low-rank approximation, matrix sketching, and principal component analysis~\cite{CohenEMMP15,CohenMM17,BravermanDMMUWZ20,MuscoM20}.

We next recall the notion of ridge leverage scores, which generalize classical leverage scores and are fundamental for defining the sampling probability in constructing PCP sketches.
\begin{definition}[$\lambda$-ridge leverage score]
For a parameter $\lambda>0$, the \emph{$\lambda$-ridge leverage score} of a row $\ba_i\in\mathbb{R}^d$ of a matrix $\bA\in\mathbb{R}^{n\times d}$ is defined as
\[\ba_i^\top(\bA^\top\bA + \lambda \bI_d)^{-1}\ba_i .\]
In our setting, we choose $\lambda_\bA := \frac{\|\bA-\bA_k\|_F^2}{k}$, where $\bA_k$ denotes the best rank-$k$ approximation to $\bA$.
\end{definition}
  
Sampling rows proportionally to their $\lambda_\bA$-ridge leverage scores produces a $(1+\eps)$-PCP sketch.  
We summarize the standard offline guarantee below.

\begin{theorem}[Offline rank-$k$ PCP sketch]
\thmlab{thm:offline:pcp}
\cite{CohenMM17}
Given a matrix $\bA\in\mathbb{R}^{n\times d}$, there is an algorithm which, with probability at least $1-\delta$, outputs a PCP sketch matrix $\bM$ consisting of
\[
f\!\left(k,\frac{1}{\eps},\frac{1}{\delta}\right)
= \mathcal{O}\!\left(\frac{k}{\eps^2}\log\frac{k}{\delta}\right)
\]
rescaled rows of $\bA$.
\end{theorem}

Similarly, we extend the ridge leverage score to the online setting.

\begin{definition}[Online $\lambda$-ridge leverage score]
For a parameter $\lambda>0$, the \emph{online $\lambda$-ridge leverage score} of a row $\ba_i\in\mathbb{R}^d$ in a matrix $\bA\in\mathbb{R}^{n\times d}$ is
\[
\ba_i^\top(\bA_{i-1}^\top\bA_{i-1} + \lambda \bI_d)^{-1}\ba_i,
\]
where $\bA_{i-1}$ denotes the prefix matrix formed by the first $i-1$ rows of $\bA$.
\end{definition}

In the streaming setting, the value $\lambda_\bA = \frac{\|\bA-\bA_k\|_F^2}{k}$ is not known in advance.  
Instead, for each incoming row $\ba_i$, one sets $\lambda_i = \Theta\!\left(\frac{\|\bM_{i-1} - (\bM_{i-1})_k\|_F^2}{k}\right)$, where $\bM_{i-1}$ consists of the previously sampled and rescaled rows.  
Because this estimate never exceeds the true $\lambda_\bA$, the resulting procedure oversamples but still guarantees valid PCP preservation.

We now state the performance guarantee of the online PCP construction.

\begin{theorem}[Online rank-$k$ PCP sketch]
\thmlab{thm:online:pcp}
\cite{BravermanDMMUWZ20}
Let $\eps>0$, $k\ge 1$, and let $\bA\in\mathbb{R}^{n\times d}$ be a matrix whose rows arrive sequentially with condition number $\kappa$.  
There exists an online algorithm which, with high probability, outputs a PCP sketch matrix $\bM$ consisting of
\[
g\!\left(n,k,\frac{1}{\eps},\kappa\right)
= \mathcal{O}\!\left(\frac{k}{\eps^2}\log n \log^2 \kappa\right)
\]
rescaled rows of $\bA$.
\end{theorem}

Finally, we recall a technical bound on the cumulative online ridge leverage scores, which controls the number of sampled rows.

\begin{lemma}[Upper bound on the sum of online ridge leverage scores]
\lemlab{lem:ub:online:ridge}
\cite{BravermanDMMUWZ20}
Let $\bA\in\mathbb{R}^{n\times d}$ have condition number $\kappa$.  
For each $i\in[n]$, define
\[
\lambda_{\ba_i}
:= \frac{\|\bA_{i-1} - (\bA_{i-1})_k\|_F^2}{k},
\qquad
\sigma(\ba_i)
:= \ba_i^\top(\bA_{i-1}^\top\bA_{i-1} + \lambda_{\ba_i}\bI_d)^{-1}\ba_i .
\]
Then the total online ridge leverage satisfies $\sum_{i=1}^n \sigma(\ba_i) = \O{k \log^2 \kappa}$.
\end{lemma}

\subsection{Efficient Encoding for Coreset Construction for Projection-Cost Preservation}
\applab{sec:pcp:encoding}



In this section, we describe the efficient encoding for PCP.
The strategy is to round a PCP sketch to its optimal rank-$k$ approximation: we first run the algorithm in \thmref{thm:offline:pcp} to obtain a $(1+\eps)$-PCP $\bB$.
Let $\bB \in \mathbb{R}^{r \times d}$.
Let $\bB_k$ be the optimal rank-$k$ approximation to $\bB$, we round $\bD = \bB - \bB_k$ using entry-wise relative error quantization with precision parameter $\eps'$ to obtain $\bD'$.
Then we store $(\bB_k, \bD')$ in the memory, and the reconstruction matrix is defined as $\bB' := \bB_k + \bD'$.
The following lemma shows that the encoding gives a $(1+\eps)$-PCP of matrix $\bB$.

\begin{lemma}
\lemlab{lem:local:encode:pcp}
For all rank $k$ orthogonal projection matrices $\bP \in \mathbb{R}^{d \times d}$,
\[(1-\eps)\|\bB-\bB \bP\|_F^2 \leq\|\bB'-\bB' \bP\|_F^2 \leq(1+\eps)\|\bB-\bB \bP\|_F^2.\]
\end{lemma}
\begin{proof}
Let $\bQ = \bI_d - \bP$, it suffices to show that
\[\left|\|\bB'\bQ\|_F^2 - \|\bB\bQ\|_F^2 \right| \le \eps \|\bB\bQ\|_F^2.\]
Now, we define the error matrix $\bE = \bB' - \bB = \bD' - \bD$, then we have
\[\|\bB'\bQ\|_F^2 = \|(\bB + \bE)\bQ\|_F^2 = \|\bB\bQ\|_F^2 + 2\langle\bB\bQ, \bE\bQ \rangle_F+\|\bE\bQ\|_F^2.\]
Then, we can decompose
\[\left|\|\bB'\bQ\|_F^2 - \|\bB\bQ\|_F^2 \right| = \left|2\langle\bB\bQ, \bE\bQ \rangle_F+\|\bE\bQ\|_F^2 \right| \le 2\|\bB\bQ\|_F \|\bE\bQ\|_F +\|\bE\bQ\|_F^2,\]
where the second step is by Cauchy-Schwartz inequality.
Since $\bQ$ is a projection matrix, we have $\|\bE \bQ\|_F \le \|\bE\|_F \|\bQ\|_2 = \|\bE\|_F$.
Moreover, the entry-wise quantization of $\bD$ gives us 
\[\|\bE\|_F = \|\bD' - \bD\|_F \le \eps' \|\bD\|_F = \eps' \|\bB -\bB_k\|_F. \]
Since $\bB_k$ is the best low-rank approximation to $\bB$ and $\bP$ is a rank-$k$ projection matrix, we have
\[\|\bB -\bB_k\|_F \le \|\bB - \bB \bP\|_F = \|\bB \bQ\|_F\]
Then we have
\[\left|\|\bB'\bQ\|_F^2 - \|\bB\bQ\|_F^2 \right| = 3\eps' \|\bB\bQ\|_F^2.\]
Taking $\eps' = \frac{\eps}{3}$ gives us the desired bound.
\end{proof}
We next upper bound the space complexity of the encoding.
\begin{lemma}
\lemlab{lem:space:local:pcp}
The efficient encoding scheme stores $\O{\left(\frac{k^2}{\eps^2} + kd \right) \log n + dr \cdot \left(\log \frac{1}{\eps} + \log\log n\right)}$ bits in memory.
In addition, reconstruction takes $\O{rdk}$ time.
\end{lemma}
\begin{proof}
We first analyze the space required to store the encoding $\bD' \in \mathbb{R}^{r \times d}$.
We use entry-wise rounding with parameter $\eps'$, e.g., by rounding it to the nearest power of $(1+\eps')$.
The number of bits required for the base is $\O{\log \frac{1}{\eps'}} = \O{\log \frac{1}{\eps}}$.
The number of bits required for the exponent depends on the range of the entries in $\bD$, which is $\poly(n)$ since we assume that the entry of $\bA$ is in $[\frac{1}{\poly(n)},\poly(n)]$, and so the bits for the exponent are $\O{\log \frac{\log n}{\eps'}} = \O{\log \frac{1}{\eps} + \log\log n}$.
Thus total bit complexity to store $\bD$ is:
\[ \O{dr \cdot \left(\log \frac{1}{\eps} + \log\log n\right)}.\]
In addition, we need to store the best rank-$k$ approximation $\bB_k$ with high precision, using $\O{\left(\frac{k^2}{\eps^2} + kd \right) \log n}$ bits, which gives the total space complexity.
Moreover, reconstruction computes $\bB' = \bB  + \bD_i'$ in $\O{rdk}$ time.
\end{proof}

Thus, we have the following theorem statement for the local encoding.
\begin{theorem}[Local encoding for PCP]
\thmlab{thm:local:encode:pcp}
Let $\bA\in\mathbb{R}^{n\times d}$ be a matrix with entries in $\{0\}\cup\left[\frac{1}{\poly(n)},\poly(n)\right]$ and let $\bB \in \mathbb{R}^{r \times d}$ be a $(1+\eps)$-PCP of $\bA$. 
Then there exists a $(1+\eps)$-PCP of $\bB$ using $\O{\left(\frac{k^2}{\eps^2} + kd \right) \log n + dr \cdot \left(\log \frac{1}{\eps} + \log\log n\right)}$ bits of space to store and $\O{rdk}$ time to reconstruct. 
\end{theorem}

\paragraph{Global efficient encoding.}

Similar to $(k,z)$-clustering and $L_p$ subspace embedding, immediately applying the previous efficient encoding to each node of a merge-and-reduce tree loses a $\polylog(\log(nd))$ factor due to the height of the merge-and-reduce tree. 
Again, we bypass this issue by rounding each local PCP to the top-$k$ right singular vectors of a global constant-factor approximation. 
We then show that the error of maintaining such a global encoding does not compound too much over the entire stream. 
We present the formal description in \algref{alg:global:efficient:encoding}.

\begin{algorithm}[!htb]
\caption{Global efficient encoding for low-rank approximation}
\alglab{alg:global:efficient:encoding}
 \begin{algorithmic}[1]
\Require{Accuracy parameter $\eps < \frac{1}{2}$, matrix $\bA = \bQ_1 \circ \ldots \circ \bQ_m$, matrix $\bB \in \mathbb{R}^{r \times d}$ that is a $(1+\eps)$-PCP of $\bA$, matrices $\bW_1,\ldots,\bW_m \in \mathbb{R}^{r \times d}$ where each $\bW_i$ is a $(1+\eps)$-PCP of $\bQ_i$, rank $k$}
\Ensure{Encodings $\bW_1', \ldots, \bW_m'$.}
\State Let $\eps'$ denote the quantization precision to be determined later. Let $\eps' \le \frac{\eps}{4 \sqrt{C_\eps}}$, where $C_\eps = \frac{1+\eps}{1-\eps}$
\State Compute the SVD of $\bB$ and let $\bV_k \in \mathbb{R}^{d \times k}$ be the top-$k$ right singular vectors. Define projections $\bP_\bB = \bV_k\bV_k^\top$ and $\bP_\bB^\perp = \bI_d - \bP_\bB$
\State Store $\bV_k$ using high precision (e.g., $\O{\log n}$ bits per entry)
\For{$i \in [m]$}
\State Compute head coefficients $\bH_i = \bW_i \bV_k$ and store with high precision
\State Compute tail residual $\bT_i = \bW_i \bP_\bB^\perp$ 
\State Round $\bT_i$ entry-wise with relative precision $\eps'$ to obtain $\bT_i'$. Store $\bT_i$
\EndFor
\State In reconstruction, compute $\bW_i' = \bH_i \bV_k^\top + \bT_i'$ for each $i \in [m]$
\State{\Return $\bW' = \bW'_1 \circ \ldots \circ \bW'_m$}
\end{algorithmic}
\end{algorithm}

First, we prove a useful lemma that bounds the Frobenius norm of the projected matrix. 
\begin{lemma}
\lemlab{lem:bound:proj}
For any projection $\bQ$, we have $\|\bW_i \bQ\|_F^2 \le C_\eps \|\bB\bQ\|_F^2$, where $C_\eps = \frac{1+\eps}{1-\eps}$.
\end{lemma}
\begin{proof}
Since $\bW_i$ is a $(1+\eps)$-PCP of $\bQ_i$, $\|\bW_i \bQ\|_F^2 \le (1+\eps)\|\bQ_i \bQ\|_F^2$.  
As $\bA$ is the concatenation of $\bQ_j$, we have
\[\|\bA \bQ\|_F^2 = \sum_{j=1}^m \|\bQ_j \bQ\|_F^2 \ge \|\bQ_i \bQ\|_F^2.\]
Now, since $\bB$ is a $(1+\eps)$-PCP of $\bA$, $\|\bB\bQ\|_F^2 \ge (1-\eps)\|\bA\bQ\|_F^2$.  
Combining these inequalities yields
\[
\|\bW_i \bQ\|_F^2 \le \frac{1+\eps}{1-\eps} \cdot \|\bB\bQ\|_F^2 = C_\eps \|\bB\bQ\|_F^2.
\]
\end{proof}

Using the previous property, we can now show that the global efficient encoding gives a PCP with $\eps$-additive error, which can subsequently be transformed into multiplicative error for our downstream purposes. 

\begin{lemma}
\lemlab{lem:global:pcp}
For all rank $k$ orthogonal projection matrices $\bP \in \mathbb{R}^{d \times d}$ and each $i \in [m]$,
\[\|\bW_i-\bW_i \bP\|_F^2 - \eps \| \bB - \bB \bP\|_F^2 \leq\|\bW_i'-\bW_i' \bP\|_F^2 \leq\|\bW_i-\bW_i \bP\|_F^2 + \eps \| \bB - \bB \bP\|_F^2.\]
\end{lemma}
\begin{proof}
Let $\bQ = \bI_d - \bP$ and fix an arbitrary $i \in [m]$.  
It suffices to show that
\[\left|\|\bW_i'\bQ\|_F^2 - \|\bW_i\bQ\|_F^2\right| \le \eps \|\bB\bQ\|_F^2.\]
Let $\bE_i = \bW_i' - \bW_i = \bT_i' - \bT_i$, then by the analysis in \lemref{lem:local:encode:pcp}, the projection-cost error satisfies
\[\left|\|\bW_i'\bQ\|_F^2 - \|\bW_i\bQ\|_F^2\right|\le 2\|\bW_i\bQ\|_F\|\bE_i\bQ\|_F + \|\bE_i\bQ\|_F^2.\]
Since $\bQ$ is a projection, $\|\bE_i\bQ\|_F \le \|\bE_i\|_F \le \eps' \|\bT_i\|_F$, and therefore,
\[\left|\|\bW_i'\bQ\|_F^2 - \|\bW_i\bQ\|_F^2\right| \le 2\eps'\|\bW_i\bQ\|_F\|\bT_i\|_F + (\eps')^2\|\bT_i\|_F^2.\]
By \lemref{lem:bound:proj}, $\|\bW_i\bQ\|_F^2 \le C_\eps \|\bB\bQ\|_F^2$.
In addition, since $\bP_\bB$ is the projection matrix onto the top-$k$ singular vectors, we have
\[\|\bT_i\|_F^2 = \|\bW_i\bP_\bB^\perp\|_F^2\le C_\eps \|\bB\bP_\bB^\perp\|_F^2 \le C_\eps \|\bB\bQ\|_F^2.\]  
Substituting these bounds gives
\[\left|\|\bW_i'\bQ\|_F^2 - \|\bW_i\bQ\|_F^2\right| \le C_\eps(2\eps' + (\eps')^2)\|\bB\bQ\|_F^2.\]
Setting $\eps' \le \frac{\eps}{4\sqrt{C_\eps}}$ gives the desired result of the global efficient encoding scheme.
\end{proof}

Unfortunately, the previous lemma is not sufficient for our purpose, as we need to show that the reconstructed matrix gives a good approximation to the online ridge leverage score of the next row in the stream.
To ensure this, we provide the following Loewner order bound of the global encoding.
This statement shows that the reconstructed sketch preserve the subspace embedding properties up to a small additive error, which suffices for our use the estimate the next online ridge leverage score.

\begin{lemma}
\lemlab{lem:global:pcp:loewner}
Let $\lambda_\bB := \frac{\|\bB - \bB_k\|_F^2}{k}$.Then the following additive Loewner bound holds for each $i\in [m]$:
\[ \frac{1}{2}\cdot \bW_i^\top \bW_i - 3 (\eps')^2 k \lambda_\bB \cdot \bI_d \preceq \bW_i'^\top \bW_i'\preceq \frac{3}{2}\cdot \bW_i^\top \bW_i + 9(\eps')^2 k \lambda_\bB \cdot \bI_d.\]
\end{lemma}

\begin{proof}
We recall that \algref{alg:global:efficient:encoding} writes $\bW_i = \bH_i \bV_k^\top + \bT_i$ and $\bW_i' = \bH_i \bV_k^\top + \bT_i'$, and we use the definition of the error matrix in \lemref{lem:global:pcp}: let $\bE_i := \bW_i' - \bW_i = \bT_i' - \bT_i$. 
Entrywise relative quantization with precision $\eps'$ ensures that $\|\bE_i\|_F \le \eps' \|\bT_i\|_F$.
Since $\bT_i = \bW_i \bP_B^\perp$ and $\bP_B$ is a projection matrix, \lemref{lem:bound:proj} gives
\[\|\bT_i\|_F^2= \|\bW_i \bP_B^\perp\|_F^2 \le  C_\eps \cdot \|\bB \bP_B^\perp\|_F^2. \]
Because $\bB \bP_B^\perp = \bB - \bB_k$, we have
\[ \|\bB \bP_B^\perp\|_F^2 = \|\bB - \bB_k\|_F^2 = k\lambda_\bB,\]
and hence $\|\bT_i\|_F^2 \le C_\eps k \lambda_\bB$.
Then we obtain
\[ \|\bE_i\|_F^2 \le (\eps')^2 \cdot \|\bT_i\|_F^2 \le(\eps')^2 C_\eps k \lambda_\bB.\]
Thus, we bound the Gram matrix of the error matrix:
\[
\bE_i^\top \bE_i 
\preceq
\|\bE_i\|_F^2 \cdot \bI_d
\preceq
(\eps')^2 C_\eps k \lambda_\bB \cdot \bI_d.
\]
Next, we prove the Loewner bound. We first expand the Gram matrix of $\bW_i'$:
\[\bW_i'^\top \bW_i'=\bW_i^\top \bW_i+ \bW_i^\top \bE_i+ \bE_i^\top \bW_i+ \bE_i^\top \bE_i.\]
For any $\eta>0$, applying operator Young’s inequality, we have
\[
-(\eta\,\bW_i^\top \bW_i + \eta^{-1}\bE_i^\top\bE_i)
\preceq
\bW_i^\top \bE_i + \bE_i^\top \bW_i
\preceq
\eta\,\bW_i^\top \bW_i + \eta^{-1}\bE_i^\top\bE_i .
\]
Taking $\eta = \tfrac{1}{2}$, since $\bE_i^\top\bE_i \preceq (\eps')^2 C_\eps k \lambda_\bB \cdot \bI_d$, we have the following upper bound:
\begin{align*}
\bW_i'^\top \bW_i'& \preceq \bW_i^\top \bW_i + \frac12 \bW_i^\top \bW_i + 2\bE_i^\top\bE_i + \bE_i^\top\bE_i \\
& \preceq \frac32\bW_i^\top\bW_i
+ 3(\eps')^2 C_\eps k \lambda_\bB \cdot \bI_d.
\end{align*}
Similarly, we have the following lower bound:
\begin{align*}
\bW_i'^\top \bW_i' & \succeq \bW_i^\top \bW_i - \left(\frac12 \bW_i^\top\bW_i + 2\bE_i^\top\bE_i\right) + \bE_i^\top\bE_i \\
& \succeq \frac12 \bW_i^\top\bW_i - (\eps')^2 C_\eps k \lambda_\bB \cdot \bI_d.
\end{align*}
Moreover since we assume that $\eps < \frac{1}{2}$, we have $C_\eps \le 3$, and then combining these two bounds gives our desired result.
\end{proof}

We now analyze the space to store the global encoding.
\begin{lemma}
\lemlab{lem:pcp:global:time}
The global efficient encoding scheme stores 
\[\O{kd\log n + mrk\log n + mrd\left(\log \frac{1}{\eps'}+ \log\log n\right)}\]
bits in memory.
In addition, the reconstruction time is $\O{mrdk}$.
\end{lemma}
\begin{proof}
First, we store $\bV_k \in \mathbb{R}^{d \times k}$ with high precision, using $\O{kd \log n}$ bits of space.
Second, we store each head coefficient $\bH_i \in \mathbb{R}^{r \times k}$ with high precision, using $\O{mrk \log n}$ bits of space.
Lastly, we store each tail residual $\bT_i'$ using entry-wise quantization with precision parameter $\eps'$.
By \lemref{lem:space:local:pcp}, the quantization stores $\O{\log \frac{1}{\eps'}+ \log\log n}$ bits per entry, and so the space usage is $\O{mrd\left(\log \frac{1}{\eps'}+ \log\log n\right)}$.
Summing up the three parts gives the total space complexity.
In addition, reconstruction computes $\bW_i' = \bH_i \bV_k^\top + \bT_i'$ in $\O{rdk}$ time for each $i \in [m]$, giving $\O{mrdk}$ overall.
\end{proof}

We now give the full guarantee of the global encoding scheme.
\begin{theorem}[Global encoding for PCP]
\thmlab{thm:global:encode:pcp}
Let $\bA = \bQ_1 \circ \ldots \circ \bQ_m\in\mathbb{R}^{n\times d}$ be a matrix with entries in $\{0\}\cup\left[\frac{1}{\poly(n)},\poly(n)\right]$, let $\bB \in \mathbb{R}^{r \times d}$ be a $(1+\eps)$-PCP of $\bA$, and let $\bW_i \in \mathbb{R}^{r \times d}$ be a $(1+\eps)$-PCP of $\bQ_i$ for each $i \in [m]$, where $\bB$ and $\bW_i$'s are obtained from ridge-leverage score sampling.
Let $\lambda_\bA := \frac{\|\bA - \bA_k\|_F^2}{k}$, there is an algorithm that encodes $\bW_i$ to $\bW'_i$ with precision $\eps' \le \O{\eps}$ for each $i \in [m]$ such that it satisfies: let $\bW = \bW'_1 \circ \ldots \circ \bW'_m$, then for all rank $k$ orthogonal projection matrices $\bP \in \mathbb{R}^{d \times d}$ and each $i \in [m]$,
\[(1- \O{m\eps}) \| \bA - \bA \bP\|_F^2 \leq\|\bW-\bW \bP\|_F^2 \leq(1+ \O{m\eps}) \| \bA - \bA \bP\|_F^2.\]
In addition, we have the Loewner bound:
\[ \frac{1}{4}\cdot \bA^\top \bA - \O{ (\eps')^2 m k \lambda_\bA} \cdot \bI_d \preceq \bW^\top \bW\preceq 2\cdot \bA^\top \bA + \O{ (\eps')^2 m k \lambda_\bA}\cdot \bI_d.\]
The algorithm uses
\[\O{kd\log n + mrk\log n + mrd\left(\log\frac{1}{\eps'} + \log\log n\right)}\]
bits of space to store and $\O{mrdk}$ time to reconstruct. 
\end{theorem}

\begin{proof}
Fix an arbitrary rank $k$ orthogonal projection matrices $\bP \in \mathbb{R}^{d \times d}$.
First, we prove the PCP property
By \lemref{lem:global:pcp}, we have for each $i \in [m]$,
\[\|\bW_i-\bW_i \bP\|_F^2 - \eps \| \bB - \bB \bP\|_F^2 \leq\|\bW_i'-\bW_i' \bP\|_F^2 \leq\|\bW_i-\bW_i \bP\|_F^2 + \eps \| \bB - \bB \bP\|_F^2.\]
Since $\bB$ is a $(1+\eps)$-PCP of $\bA$, we have $\| \bB - \bB \bP\|_F^2 = \Theta(1) \| \bA - \bB \bA\|_F^2$, which gives
\[\|\bW_i-\bW_i \bP\|_F^2 - \O{\eps} \| \bA - \bA \bP\|_F^2 \leq\|\bW_i'-\bW_i' \bP\|_F^2 \leq\|\bW_i-\bW_i \bP\|_F^2 + \O{\eps} \| \bA - \bA \bP\|_F^2.\]
Moreover, since each $\bW_i$ is a $(1+\eps)$-PCP of $\bQ_i$, we have 
\begin{align*}
& |\bW_i'-\bW_i' \bP\|_F^2 \ge (1-\eps) \|\bQ_i-\bQ_i \bP\|_F^2 - \O{\eps} \| \bA - \bA \bP\|_F^2\\
&|\bW_i'-\bW_i' \bP\|_F^2 \leq (1+\eps) \|\bQ_i-\bQ_i \bP\|_F^2 + \O{\eps} \| \bA - \bA \bP\|_F^2.
\end{align*}
Now, because $\bA = \bQ_1 \circ \ldots \circ \bQ_m$ we have $\| \bA - \bA \bP\|_F^2 = \sum_{i \in [m]} \|\bQ_i-\bQ_i \bP\|_F^2$, and the same holds for $\bW = \bW'_1 \circ \ldots \circ \bW'_m$.
Therefore, summing across all $i \in [m]$, we have
\[(1- \O{m\eps}) \| \bA - \bA \bP\|_F^2 \leq\|\bW-\bW \bP\|_F^2 \leq(1+ \O{m\eps}) \| \bA - \bA \bP\|_F^2.\]
Next, we show the Loewner bound. From \lemref{lem:global:pcp:loewner}, we have for each $i\in [m]$,
\[ \frac{1}{2}\cdot \bW_i^\top \bW_i - \O{ (\eps')^2 k \lambda_\bB} \cdot \bI_d \preceq \bW_i'^\top \bW_i'\preceq \frac{3}{2}\cdot \bW_i^\top \bW_i + \O{(\eps')^2 k \lambda_\bB} \cdot \bI_d,\]
where $\lambda_\bB := \frac{\|\bB - \bB_k\|_F}{k} = \Theta(\lambda_\bA)$.
Thus,
\[ \frac{1}{2}\cdot \bW_i^\top \bW_i - \O{ (\eps')^2 k \lambda_\bA} \cdot \bI_d \preceq \bW_i'^\top \bW_i'\preceq \frac{3}{2}\cdot \bW_i^\top \bW_i + \O{(\eps')^2 k \lambda_\bA} \cdot \bI_d.\]
Note that $\bW_i$ is obtained from ridge-leverage score sampling, then we have the following Loewner order property (c.f., Theorem 5 in \cite{CohenMM17})
\[(1-\eps)\bQ_i^\top \bQ_i -\eps \lambda_\bA \cdot \bI_d \preceq \bW_i^\top \bW_i \preceq (1+\eps)\bQ_i^\top \bQ_i +\eps \lambda_\bA \cdot \bI_d.\]
Therefore, we have
\[ \frac{1}{4}\cdot \bQ_i^\top \bQ_i - \O{ (\eps')^2 k \lambda_\bA} \cdot \bI_d \preceq \bW_i'^\top \bW_i'\preceq 2\cdot \bW_i^\top \bW_i + \O{(\eps')^2 k \lambda_\bA} \cdot \bI_d.\]
Again, since $ \bA^\top \bA = \sum_{i \in [m]} \bQ_i ^\top \bQ_i$, summing across each $i\in[m]$, we have
\[ \frac{1}{4}\cdot \bA_i^\top \bA_i - \O{ (\eps')^2 m k \lambda_\bA} \cdot \bI_d \preceq \bW^\top \bW\preceq 2\cdot \bW_i^\top \bW_i + \O{ (\eps')^2 m k \lambda_\bA}\cdot \bI_d.\]
Lastly, the space usage and the reconstruction time follow from \lemref{lem:pcp:global:time}.
\end{proof}

\subsection{Projection-Cost Preservation in the Streaming Model}
\applab{sec:pcp:space}

In this section, we show how to use the global encoding from the previous section to achieve our streaming algorithm for projection-cost preservation with optimal space.
We give the algorithm in full in \algref{alg:pcp:insert:stream}, which applies our space-efficient framework in \algref{alg:framework:space}. 

\begin{algorithm}[!htb]
\caption{Projection-Cost preservation on insertion-only streams}
\alglab{alg:pcp:insert:stream}
\begin{algorithmic}[1]
\Require{Matrix $\bA=\{\ba_1,\ldots,\ba_n\}\in\mathbb{R}^{n\times d}$ that arrives as a data stream, $\eps\in(0,1)$, parameter $p\ge 1$}
\Ensure{$(1+\eps)$-coreset for projection-cost preservation}
\State{$\bZ\gets\emptyset$, $\alpha\gets\O{\frac{\log n}{\eps^2}}$}
\For{each $t\in[n]$}
\State{Use $\bZ\cup\{\widehat{\ba_t}\}$ to compute the online $\lambda$-ridge leverage score $\tau(\ba_t)$ of $\ba_t$, where $\lambda = \frac{\|\bZ - \bZ_k\|_F^2}{k}$}
\State{$p(\ba_t)\gets\min(1,\alpha\cdot\tau(\ba_t))$}
\State{With probability $p(\ba_t)$, sample $\frac{1}{p(\ba_t)}\cdot\ba_t$ into a stream $\calS'$}
\State{Let $\bZ$ be the running output of merge-and-reduce on $\calS'$ using the global encoding for coreset in \algref{alg:global:efficient:encoding}, for accuracy $\frac{\eps}{\poly(\log\log n \kappa)}$, failure probability $\frac{1}{\poly\left(\frac{1}{\eps},\log(n\kappa)\right)}$, and precision $\eps' = \frac{1}{\sqrt{k}} \cdot \poly(\log k, \log \frac{1}{\eps}, \log \log n)$} \Comment{Detailed implementation of \algref{alg:framework:space}}
\EndFor
\State{\Return $\bZ$}
\end{algorithmic}
\end{algorithm}

We now give the full guarantees of \algref{alg:pcp:insert:stream}.
We remark that our update time in each step is dominated by reconstructing the sketch and computing the online sampling probability, and so the update time is $\poly(d)$, where the power of $d$ depends on the matrix-multiplication exponential.

\begin{theorem}
\thmlab{thm:pcp:per:update}
Given a matrix $\bA$ of $n$ rows on $[-M,\ldots,-1,0,1,\ldots,M]^d$, with $M=\poly(n)$ and online condition number $\kappa$, there exists a one-pass streaming algorithm in the row arrival model that outputs a $(1+\eps)$-coreset of $\bA$ and uses $\O{k \polylog\left(\frac{1}{\eps},\log(nd\kappa)\right) +d}\cdot f\left(d,\frac{1}{\eps},\frac{1}{0.01}\right)$ words of space, where $f\left(k, \frac{1}{\eps}, \frac{1}{\delta}\right) = \mathcal{O}\left(\frac{k}{\eps^2} \log \frac{k}{\delta}\right)$.
The (amortized) update time of this algorithm is $\poly(d)$.
\end{theorem}
\begin{proof}
The proof mainly follows from the proof of $L_p$ subspace embedding in \thmref{thm:subspace:per:update}.
However, the main difference is that the PCP property does not necessarily imply the subspace embedding property, which is essential in showing the validity of the online sampling probability.
Thus, we need to use a finer precision parameter $\eps'$ and apply the Loewner order bound in \thmref{thm:global:encode:pcp}.

Consider \algref{alg:pcp:insert:stream}. 
Again, we prove correctness by induction on $t$, claiming that after $j\cdot f\left(k,\frac{1}{\eps},\frac{1}{\delta}\right)$ samples into $\calS'$, then $\bZ$ is a $\left(1+\O{\eps}\right)^{j+1}$ coreset of $\bA_t$.
The online ridge leverage score of the first row is always $1$ and thus $\bZ=\ba_1$ after the first step, and the base case is complete. 
For the inductive hypothesis, we condition on the correctness of $\bZ$ being a $\left(1+\O{\eps}\right)^j$-coreset of $\bA_{t-1}=\ba_1\circ\ldots\circ\ba_{t-1}$ after $j-1$ merges have occured, such that $\lambda_\bZ$ is within the range of $(1\pm\O{j\eps}) \cdot \lambda_{\bA_{t-1}}$. 
Moreover, we condition on that $\bZ$ is a constant-factor subspace embedding of $\bA_{t-1}$.
Therefore, we can use $\bZ$, $\lambda_\bz$, and $\ba_t$ to compute a constant-factor approximation to the desired online $\lambda$-ridge leverage score $\ba_t^\top(\bA_{t-1}^\top\bA_{t-1} + \lambda_{\bA_t} \cdot \bI_d)^{-1}\ba_t$, which suffices to give a valid sampling probability.
Then by \thmref{thm:online:pcp} through \lemref{lem:ub:online:ridge}, sub-stream $\calS'$ has $g \left(n,k,\frac{1}{\eps}, \kappa \right) = \mathcal{O}\left(\frac{k}{\eps^2} \log n \log ^2 \kappa\right)$ edges.
Note that merge-and-reduce will consider at least $f\left(d,\frac{1}{\eps},\frac{1}{\delta}\right)$ stream updates before applying a new coreset construction and the failure probability is set to $\frac{1}{\poly\left(\frac{1}{\eps}, \log(n \kappa)\right)}$, so that $j\le \O{\log \frac{g\left(n,d,\frac{1}{\eps},\kappa, p\right)}{f\left(d,\frac{1}{\eps},p\right)}}\le\poly(\log \log(n\kappa))$ and the failure probability can be union bound over the $j$ merges. 
Thus by \thmref{thm:global:encode:pcp}, we have that $\bZ$ after time $t$ will be a $\left(1+\O{\eps}\right)^{j+1}$-strong coreset for $\bA_t$ after $j$ merges; and moreover, by our choice of precision $\eps' = \frac{1}{\sqrt{k}} \cdot \poly(\log k, \log \frac{1}{\eps}, \log \log n)$ in \thmref{thm:global:encode:pcp}, $\bZ$ still gives a constant-factor subspace embedding for $\bA_t$, which completes the induction.
The correctness guarantee then follows by rescaling $\eps$ with $\poly(\log \log(n\kappa))$. 

To analyze the space complexity, note that each coreset in the merge-and-reduce tree has size $f\left(d,\frac{\poly(\log \log(n\kappa))}{\eps},\poly\left(\frac{1}{\eps}, \log(n\kappa)\right)\right)$. 
We use the global encoding for coreset construction given in \algref{alg:global:efficient:encoding} with accuracy $\frac{\eps}{\poly(\log\log(n\kappa))}$, failure probability $\frac{1}{\poly\left(\frac{1}{\eps},\log(n\kappa)\right)}$, and precision parameter $\eps' = \frac{1}{\sqrt{k}} \cdot \poly(\log k, \log \frac{1}{\eps}, \log \log n)$. 
By \thmref{thm:global:encode:pcp}, the global encoding takes
\begin{align*}
& \O{kd\log n + f\left(d,\frac{\poly(\log \log(n\kappa))}{\eps},\poly\left(\frac{1}{\eps}, \log(n\kappa)\right)\right) \cdot jk\log n} + \\
& \O{f\left(d,\frac{\poly(\log \log(n\kappa))}{\eps},\poly\left(\frac{1}{\eps}, \log(n\kappa)\right)\right) \cdot jd\left(\log\frac{1}{\eps'} + \log\log n\right)}
\end{align*}
bits of space to store.
Since the height of the merge-and-reduce tree is at most $\poly(\log \log(n\kappa))$, it follows that there are $j \le \poly(\log \log(n\kappa))$ such coresets. 
In addition, by \thmref{thm:offline:pcp} we have $f\left(k, \frac{1}{\eps}, \frac{1}{\delta}\right) = \mathcal{O}\left(\frac{k}{\eps^2} \log \frac{k}{\delta}\right)$, which scales polynomially with $\frac{1}{\eps}$ and $\log \frac{1}{\delta}$.
Therefore, the total space required is 
\[\O{kd\log n} + (k \log n + d)\cdot f\left(d,\frac{1}{\eps}, \frac{1}{0.01}\right) \cdot \polylog\left(\frac{1}{\eps},\log(nd\kappa)\right).\]
in bits, which is equivalent to $\O{k \polylog\left(\frac{1}{\eps},\log(nd\kappa)\right) +d}\cdot f\left(d,\frac{1}{\eps},\frac{1}{0.01}\right)$ words of space, as desired.

To analyze the update time, in each step, we reconstruct the PCP sketch $\bZ$ from our global encoding using $\poly\left(k, \frac{1}{\eps}, \log \log(n\kappa)\right)$ update time by \lemref{lem:pcp:global:time}.
Note that we can instead store a constant-factor PCP sketch to define the online sampling probability, which suffices to give a constant-approximation to the ridge leverage score, so we can improve the update time from $\poly\left(k, \frac{1}{\eps}, \log \log(n\kappa)\right)$ to $\poly\left(k, \log \log(n\kappa)\right)$.
Then we apply $\bZ$ to estimate the online ridge leverage score of row $\ba_t$, using $\poly(d)$ time \cite{CohenMP20}.
Similar to $L_p$ subspace embedding, whenever we merge some coreset, we need $\poly\left(d, \frac{1}{\eps}, \log\log (n\kappa)\right)$ update time to compute the global efficient encoding.
However, since we only conduct $\polylog(n\kappa)$ times such operation in the entire stream, they are amortized to lower order terms and hence are dominated by the time to compute the online ridge leverage scores.
\end{proof}



\bibliography{references}
\bibliographystyle{alpha}

\appendix

\section{Experimental Results}
\seclab{sec:exp}

In this section, we provide empirical validation of our novel techniques. \footnote{The Python implementation can be found on \url{https://github.com/XieShenghao16/efficient_encoding.git}.}
Our goal is to demonstrate that our efficient encoding approach not only provides theoretical space guarantees but also yields practical improvements in memory efficiency without sacrificing approximation quality.
The experiments are conducted using an Apple M2 CPU, with 16 GB RAM and 8 cores.

\paragraph{Real-world dataset.}
We conducted our $k$-means clustering algorithms on the \textsc{MNIST} dataset \cite{lecun1998gradient}, which consists of $n=70{,}000$ images of handwritten digits $0$--$9$, and each image has a dimension of $28 \times 28$, represented by $d=784$ features corresponding to pixel intensities.
We set $k = 10$, corresponding to the ten digit classes.
\textsc{MNIST} serves as a large-scale dataset to evaluate the memory efficiency in streaming coreset construction.
The dataset is publicly available through the MNIST website and the UCI repository.

\paragraph{Baseline.}
Our primary baseline is the $k$-means coreset construction algorithm through the framework in \cite{Cohen-AddadWZ23}, where the algorithm first conducts online sensitivity sampling on the data stream, and then passes the selected points to a merge-and-reduce structure.
This baseline achieves the state-of-the-art worst-case upper bounds on space complexity among prior streaming algorithms (see \figref{fig:space:summary}).
We compare our algorithm in \figref{fig:outline}, which efficiently encodes each coreset obtained by the merge-and-reduce structure, with the baseline and demonstrate our ability in space reduction.

\paragraph{Evaluation Metrics.}
We compare the performance of the two methods on \emph{clustering cost} and \emph{space consumption}.
Once the algorithms return the $k$-means coresets at the end of the stream, we run the same $k$-means heuristic in \textsc{sklearn.cluster} on the reported coresets and calculate the clustering cost with respect to the entire dataset, which is the main evaluation of the quality of the constructed coresets.
We then compute the peak space consumption (in bits) of the two methods required to maintain the data structure during the stream.
We vary the sampling coefficients of the two methods to obtain target sample budgets of approximately $20$, $50$, $100$, and $200$, and compare the three metrics under each budget.

\paragraph{Results and Discussion.}
We summarize our empirical results on the \textsc{MNIST} dataset in \figref{fig:results}, which plots the average clustering costs and peak space consumption over $5$ repetitions.
The full result table is shown in \tableref{tab:mnist-results}.
We also include the amortized time, which is the total update time divided by the number of images.

\begin{figure}[!ht]
    \centering
    \begin{subfigure}[b]{0.475\textwidth}
        \centering
        \includegraphics[width=\textwidth]{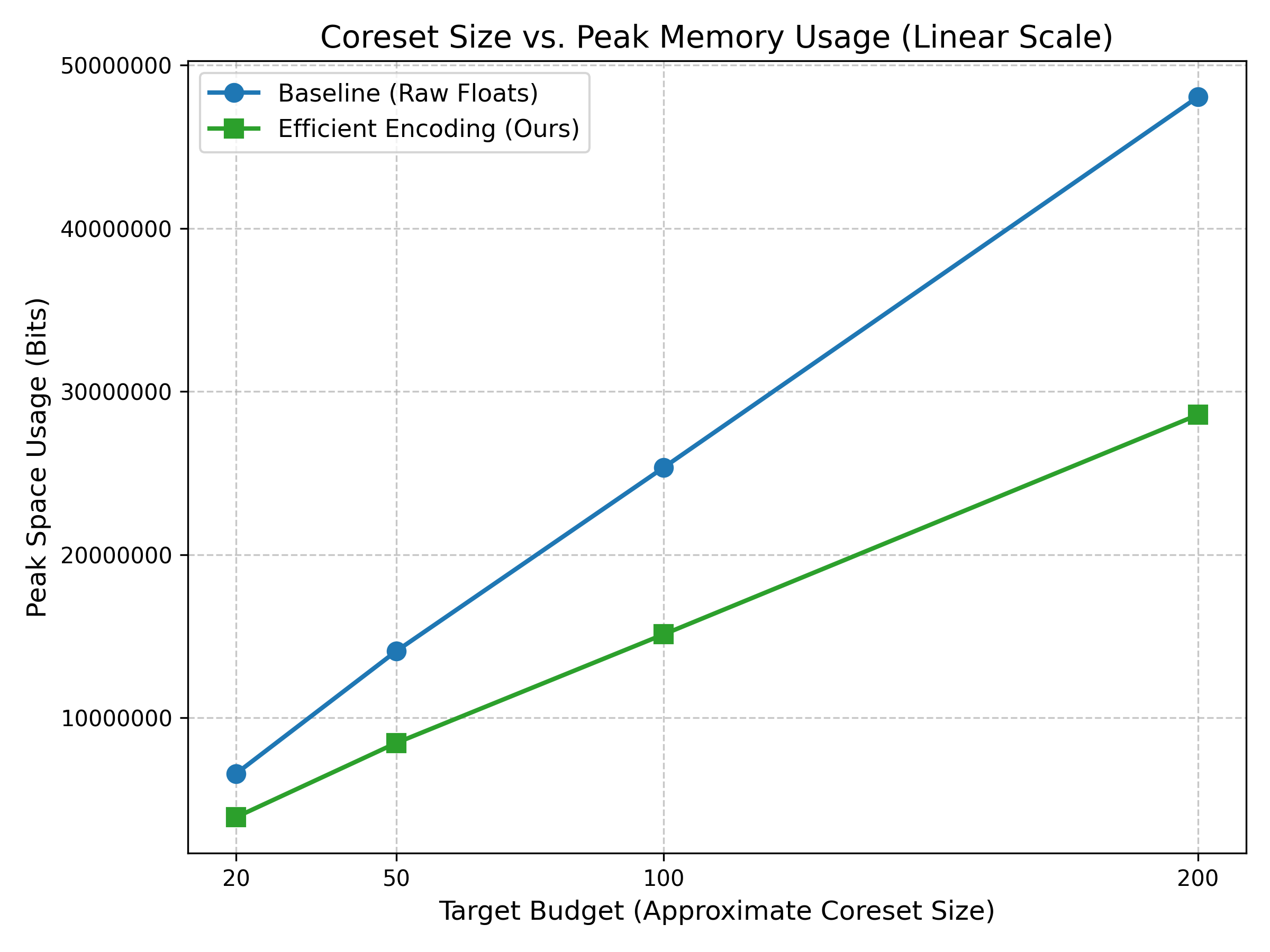}
        \caption{Space consumption vs. target budget}
    \end{subfigure}
    \hfill
    \begin{subfigure}[b]{0.475\textwidth}
        \centering
        \includegraphics[width=\textwidth]{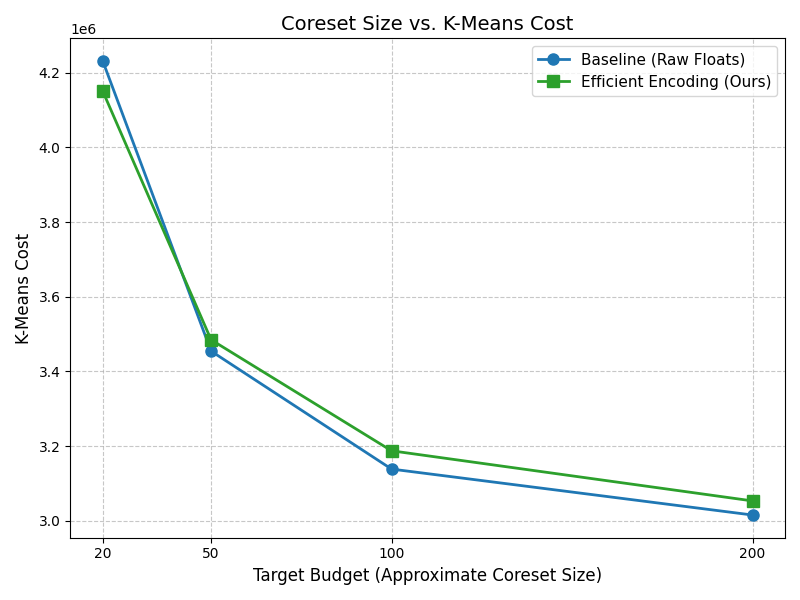}
        \caption{Clustering cost vs. target budget}
    \end{subfigure}
    \caption{Empirical comparison on the \textsc{MNIST} dataset. Target budgets indicate approximate final coreset sizes.}
    \figlab{fig:results}
\end{figure}

\begin{table}[!ht]
    \centering
    \small
    \label{tab:mnist-results}
    \begin{tabular}{@{}rlrrr@{}}
        \hline
        Target budget & Method & Mean cost & Peak space (bits) & Time/image (ms) \\
        \hline
        $20$  & Baseline & 4,231,656 &  6,581,440 &  7.15 \\
              & Ours     & 4,150,784 &  3,911,787 &  8.77 \\
        $50$  & Baseline & 3,454,396 & 14,092,320 & 10.26 \\
              & Ours     & 3,484,771 &  8,462,245 & 15.66 \\
        $100$ & Baseline & 3,138,230 & 25,337,707 & 13.40 \\
              & Ours     & 3,187,218 & 15,114,096 & 24.41 \\
        $200$ & Baseline & 3,015,100 & 48,054,560 & 14.91 \\
              & Ours     & 3,052,912 & 28,572,587 & 32.61 \\
        \hline
    \end{tabular}
    \vspace{+0.1cm}
    \caption{Clustering cost, peak space, and amortized update time on \textsc{MNIST}.}
    \tablelab{tab:mnist-results}
\end{table}

As shown in the ``Space consumption vs. target budget'' comparison in \figref{fig:results}, our efficient encoding scheme consistently outperforms the baseline on space usage, achieving a reduction of approximately $40\%$ across all tested budgets.
Thus by storing each sampled point using efficient quantization instead of high precision, our algorithm requires significantly fewer bits to store the data structure than the baseline.
Despite the reduced space, as shown in the ``Clustering cost vs. target budget'' comparison in \figref{fig:results}, our algorithm achieves clustering costs within $2\%$ of the baseline across all tested budgets, indicating that efficient encoding and reconstruction preserve comparable empirical coreset quality.
These results demonstrate the effectiveness of our efficient encoding framework in reducing memory consumption while maintaining comparable clustering quality.

\end{document}